\documentclass[twocolumn,nofootinbib,superscriptaddress,preprintnumbers,floatfix]{revtex4-1}

\usepackage{amsmath}
\usepackage{graphicx}
\usepackage[small]{subfigure}
\usepackage[hyperfootnotes=false]{hyperref} 
\usepackage{braket}
\usepackage{cancel}
\usepackage{xspace}
\usepackage{fmtcount}

\usepackage{color}
\usepackage{multirow}

\newcommand{\eq}[1]{Eq.~\eqref{eq:#1}}
\newcommand{\eqs}[2]{Eqs.~\eqref{eq:#1} and \eqref{eq:#2}}
\newcommand{\fig}[1]{Fig.~\ref{fig:#1}}
\newcommand{\tab}[1]{Tab.~\ref{tab:#1}}

\newcommand{\df}{\mathrm{d}}

\newcommand{\nn}{\nonumber}

\newcommand{\ang}[1]{\tau_{(#1)}}

\newcommand{\vev}[1]{\left\langle #1 \right\rangle}
\newcommand{\GeV}{\text{GeV}}
\newcommand{\pert}{\text{pert}}

\allowdisplaybreaks[4]

\DeclareRobustCommand{\Sec}[1]{Sec.~\ref{#1}}
\DeclareRobustCommand{\Secs}[2]{Secs.~\ref{#1} and \ref{#2}}
\DeclareRobustCommand{\App}[1]{App.~\ref{#1}}
\DeclareRobustCommand{\Tab}[1]{Table~\ref{#1}}

\DeclareRobustCommand{\Fig}[1]{Fig.~\ref{#1}}
\DeclareRobustCommand{\Figs}[2]{Figs.~\ref{#1} and \ref{#2}}
\DeclareRobustCommand{\Eq}[1]{Eq.~(\ref{#1})}
\DeclareRobustCommand{\Eqs}[2]{Eqs.~(\ref{#1}) and (\ref{#2})}
\DeclareRobustCommand{\Ref}[1]{Ref.~\cite{#1}}
\DeclareRobustCommand{\Refs}[1]{Refs.~\cite{#1}}

\newcommand{\be}{\begin{equation}}
\newcommand{\ee}{\end{equation}}

\newcommand{\Oop}{\Omega_1(r)}
\newcommand{\Eop}{\hat{\mathcal{E}}_T}

\begin{document}


\preprint{\vbox{\hbox{IFIC/12-66}\hbox{LPN12-099}\hbox{MIT--CTP 4394}
}}

\title{Power Corrections to Event Shapes with Mass-Dependent Operators}

\author{Vicent Mateu} 
\affiliation{Center for Theoretical Physics, Massachusetts Institute of
  Technology, Cambridge, MA 02139, USA}
\affiliation{IFIC, UVEG - CSIC, Apartado de Correos 22085, E-46071, 
        Valencia, Spain}
  
\author{Iain W.~Stewart} 
\affiliation{Center for Theoretical Physics, Massachusetts Institute of
  Technology, Cambridge, MA 02139, USA}

\author{Jesse Thaler\vspace{0.2cm}}
\affiliation{Center for Theoretical Physics, Massachusetts Institute of
  Technology, Cambridge, MA 02139, USA}

\begin{abstract}
  We introduce an operator depending on the ``transverse velocity'' $r$ that
  describes the effect of hadron masses on the leading $1/Q$ power correction to
  event-shape observables.  Here, $Q$ is the scale of the hard collision. This
  work builds on earlier studies of mass effects by Salam and
  Wicke~\cite{Salam:2001bd} and of operators by Lee and
  Sterman~\cite{Lee:2006nr}.  Despite the fact that different event shapes have
  different hadron mass dependence, we provide a simple method to identify
  universality classes of event shapes whose power corrections depend on a
  common nonperturbative parameter.  We also develop an operator basis to show
  that at a fixed value of $Q$, the power corrections for many classic
  observables can be determined by two independent nonperturbative matrix
  elements at the 10\% level.  We compute the anomalous dimension of the
  transverse velocity operator, which is multiplicative in $r$ and causes the
  power correction to exhibit non-trivial dependence on $Q$.  The existence of
  universality classes and the relevance of anomalous dimensions are reproduced
  by the hadronization models in Pythia~8 and Herwig++, though the two programs
  differ in the values of their low-energy matrix elements.
    
\end{abstract}

\maketitle

\section{Introduction}
\label{sec:intro}

Event shapes have played a key role in establishing the structure of quantum
chromodynamics (QCD).  Indeed, the jet-like behavior of QCD collisions at high
energies was established in 1975 by measuring the event shape sphericity in $e^+
e^-$ collisions \cite{Hanson:1975fe}.  Event shapes are used for precision
determinations of the strong coupling constant $\alpha_s$ using $e^+ e^-$ data
\cite{Bethke:2009jm}, and the same event shapes are used to tune Monte Carlo
hadronization models to help make particle-level predictions for the Large
Hadron Collider (LHC) (see e.g.~\Ref{Buckley:2009bj}).  Recently there has also been a resurgence of
interest in event shapes because they are closely related to jet shapes, which
are a sensitive probe of jet substructure
\cite{Abdesselam:2010pt,Altheimer:2012mn}.

Event shapes require both perturbative and nonperturbative contributions from
QCD.  Therefore power corrections are an important theoretical ingredient for
making event shape predictions.  In the kinematic tail region, the leading
nonperturbative effect is simply to shift perturbative event shape distributions
by an amount suppressed by $1/Q$ \cite{Dokshitzer:1997ew}.  Such $1/Q^n$ effects
are known as power corrections, where $Q$ is the scale of the hard collision.
Some power corrections are known to exhibit universality, in the sense that the
same nonperturbative shift parameter describes more than one event shape.  Power
corrections also encode the effect of hadron masses on event shape distributions
\cite{Salam:2001bd}, since different methods for treating the energy/momentum of
soft hadrons with mass $m_H$ can change event shapes by
$\mathcal{O}(m_H/Q)$.

In this paper, we revisit the theoretical underpinnings of power corrections for
$e^+ e^-$ event shapes, with a particular focus on universality and hadron
masses.  Building on the work of Salam and Wicke~\cite{Salam:2001bd}, we show
that hadron mass effects break traditional power correction universality, but we
are still able to define universality classes of event shapes with a common
power correction.  Building on the work of Lee and Sterman~\cite{Lee:2006nr}, we
provide an operator definition of the leading power correction that demonstrates
that its hadron mass dependence is described by a ``transverse velocity''
distribution.  Transverse velocity $r$ is defined as
\begin{align}
  r \equiv \frac{p_\perp}{\sqrt{p_\perp^2+m_H^2}} \,,
\end{align}
where the transverse momentum $p_\perp$ is measured with respect to the thrust
axis.  The transverse velocity operator has nontrivial renormalization group
evolution which is multiplicative in $r$, such that the leading power correction
includes a term behaving as $(\alpha_s\ln Q)/Q$.

To put our work in context, it is worthwhile to review some of the key
literature on event shapes and power corrections.  Examples of classic $e^+ e^-$
dijet event shapes are thrust~\cite{Farhi:1977sg}, the
C-parameter~\cite{Parisi:1978eg,Donoghue:1979vi}, hemisphere jet
masses~\cite{Clavelli:1979md, Chandramohan:1980ry, Clavelli:1981yh}, jet
broadening~\cite{Rakow:1981qn}, and angularities~\cite{Berger:2003iw}, all of
which were measured at LEP.  The theoretical understanding of event shapes has
undergone substantial progress in recent years, especially for perturbative
calculations.  Fixed-order corrections to event-shape distributions have been
calculated up to ${\cal O}(\alpha_s^3)$
precision~\cite{Ellis:1980wv,Catani:1996jh,Catani:1996vz,GehrmannDeRidder:2007bj,
  GehrmannDeRidder:2007hr,Weinzierl:2008iv,Weinzierl:2009ms}, and large singular
logs have been resummed to N${}^3$LL accuracy for thrust \cite{Becher:2008cf}
and heavy jet mass \cite{Chien:2010kc}, and to Next-to-Leading Logarithm for jet
broadening~\cite{Chiu:2011qc,Becher:2011pf,Chiu:2012ir}. For certain event
shapes $e$, it can be demonstrated that the leading power correction parameter
$\Omega_1^e$ appears both in the mean value of the event shapes as well as in
the tail region of the full event shape distribution~\cite{Dokshitzer:1997ew}.
This fact was discussed using a factorization theorem for thrust in
\Ref{Abbate:2012jh}, and used to simultaneously extract $\alpha_s(m_Z)$ and
$\Omega_1^\tau$ from thrust data at N${}^3$LL$\,+\,{\cal O}(\alpha_s^3)$ in
\Refs{Abbate:2010xh,Abbate:2012jh}. 

It is worth noting that heavy hadrons and heavy quark masses do not play a
direct role in the leading ${\cal O}(m_H/Q)$ power corrections for dijets.  Hadrons containing a
heavy quark decay before reaching the detector, so $m_H$ here refers to only
light hadrons. At the partonic level it is straightforward to include the
effect of heavy quark masses analytically \cite{Bernreuther:1997jn,Rodrigo:1999qg,
Brandenburg:1997pu}.  Up to the one-loop level in the
corresponding factorization theorem it is known that only the kinematic
threshold and jet function are modified~\cite{Fleming:2007qr,Fleming:2007xt},
plus smaller contributions from non-singular terms~\cite{Abbate:2010xh}.

To study nonperturbative power corrections, there have been two broad
strategies.  The first strategy is to build analytic models to describe the
nonperturbative physics.  For example, in the dispersive approach
\cite{Dokshitzer:1995zt,Dokshitzer:1995qm,Dokshitzer:1998pt}, one introduces an
infrared cutoff $\mu_I$ below which the strong coupling constant is replaced by
an effective coupling $\alpha_{\rm eff}$.  Perturbative infrared effects coming
from scales below the cutoff are subtracted, and nonperturbative effects are
parametrized in terms of an average value for the effective coupling $\alpha_0$
times $\mu_I$.\footnote{The dispersive approach was motivated by the study of
  renormalons (see \Ref{Beneke:1998ui} for a review).  Renormalons refer to an
  ambiguity in a resummed perturbative series of order $(\Lambda/Q)^p$, and this
  ambiguity is related to the nonperturbative power correction.
  Renormalon-based models were originally applied to the mean values of the
  event-shape distributions
  \cite{Dokshitzer:1995zt,Akhoury:1995sp,Akhoury:1995fb,Nason:1995hd} (see also
  \cite{Beneke:1996xy}) and later generalized to the dijet limit of
  distributions as well \cite{Korchemsky:1994is,Dokshitzer:1997ew}.}  The
original dispersive approach relied on the single-gluon approximation. In
\Refs{Dokshitzer:1997iz,Dokshitzer:1998pt} the Milan factor was
introduced as a refinement to handle multi-gluon diagrams.  Another
renormalon-inspired analytic model is provided in the dressed gluon
approach~\cite{Gardi:1999dq,Gardi:2001ny,Gardi:2002bg}.

A key prediction that emerged from these analytic models (first seen in the
dispersive approach) was universality of power corrections
\cite{Dokshitzer:1995zt,Akhoury:1995sp} (see also \cite{Korchemsky:1994is}).
Universality posits that the leading power correction $\Omega_1^e$ to an event
shape $e$ can be separated into two pieces, a calculable coefficient $c_e$ which
depends on the event shape in question times a universal nonperturbative
parameter $\Lambda$ common to all event shape observables:
\be
\Omega_1^e \to c_e \Lambda.
\ee
Because of universality, one can extract the power correction $\Lambda$ from one
event shape measurement and apply it to another one, as studied
in~\Refs{Heister:2003aj,Bauer:2002ie,Achard:2004sv,Dokshitzer:1997iz,Dokshitzer:1998pt,Korchemsky:2000kp,Gardi:2002bg,Lee:2006nr,Gehrmann:2009eh}.

The second strategy to understand power corrections
is based on QCD factorization, which implements a separation of perturbative and
nonperturbative contributions.  The shape function was introduced in
\Refs{Korchemsky:1999kt,Korchemsky:2000kp} to describe nonperturbative
corrections, and it accounts for a whole range of power corrections of the order
$(\Lambda_{\rm QCD}/eQ)^n$.  In the tail region where $\Lambda_{\rm QCD} \ll
Q\,e \ll Q$, the shape function can be expanded in terms of derivatives of the
Dirac delta function, and this translates into an operator expansion for the
event-shape distribution. This shape function can be derived in the
Collins-Soper-Sterman (CSS) approach to factorization \cite{Collins:1981uk,
  Korchemsky:1998ev,Korchemsky:1999kt,Korchemsky:2000kp,Berger:2003iw}. The
shape function also emerges naturally from factorization properties of
Soft-Collinear Effective Theory (SCET)~\cite{Bauer:2000ew, Bauer:2000yr,
  Bauer:2001ct, Bauer:2001yt, Bauer:2002nz}. Here methods exist to
systematically improve the description of the shape
function~\cite{Hoang:2007vb,Ligeti:2008ac}, and to sum large logs present in the
subtractions needed to define the power corrections in a scheme that is free from
the leading renormalon ambiguity~\cite{Hoang:2008yj,Hoang:2009yr}.

The key advantage of the factorization-based strategy is that one can express
nonperturbative parameters in terms of matrix elements involving QCD fields.
For example, the leading power correction $\Omega_1$ can be expressed in terms
of the energy flow
operator~\cite{Sveshnikov:1995vi,Korchemsky:1997sy,Lee:2006nr,Bauer:2008dt}.
Using this fact, Lee and Sterman~\cite{Lee:2006nr} rigorously derived
universality for the leading power correction without relying on an analytic
model.  This proof of power correction universality solely uses QCD first
principles.

The drawback of both of the two above strategies is that they start by making
the assumption that all final state hadrons can be considered to be massless. (An
important exception to this is \Ref{Beneke:1996xy,Beneke:1998ui}, where mass effects were
studied with renormalons using massive gluons.)  The massless assumption is certainly valid
at the parton level, but the actual hadrons measured in an event shape are of course massive.
One might erroneously argue that at very high energies the hadron masses can be
safely neglected, but because nonperturbative effects are caused by soft
particles whose momenta are of order $\Lambda_{\rm QCD} \sim m_{H}$, their masses
contribute at order $m_H/Q$, which is the same order as the leading power
correction.  One could try to study the effect of hadron masses by using Monte
Carlo hadronization models, but this approach is not fully satisfactory since
there is no clean separation between perturbative parton shower evolution and
nonperturbative hadronization effects, and there is no guarantee that one can
systematically improve the accuracy of the treatment of hadronization effects.

The first serious study of hadron mass effects on power corrections was
performed by Salam and Wicke \cite{Salam:2001bd}, where they argued that
universality does not hold for event shapes such as thrust, jet masses, or the
C-parameter with their traditional definitions.  However, they showed that
universality can be restored if one makes measurements in the E-scheme, where
one performs the following substitution in the event shape definition:
\begin{equation}\label{eq:E-scheme}
\vec{p}_i \to \frac{E_i}{|\vec{p}_i|}\,\vec{p}_i \,.
\end{equation}
\Ref{Salam:2001bd} also argues that hadron mass effects, in addition to
breaking universality, generate a power correction of the form $(\ln
Q)^A/Q$, with $A\sim 1.5$.

In this paper we will devise a rigorous operator-based method of treating hadron
mass effects in the leading event shape power correction by defining a
transverse velocity operator.  We generalize $\Omega_1^e$ to a function
$\Omega_1(r,\mu)$ which accounts for both the transverse velocity dependence
through $r$ as well as renormalization group evolution through $\mu$. Using this
operator we derive universality classes for event shapes in the presence of
hadron masses, following the treatment of \Ref{Lee:2006nr}, and show that
each event shape belongs to a unique class.  From studying the $r$- and
$\mu$-dependence, we will largely confirm the results of Salam and Wicke.

\addtocounter{footnote}{1}
\begin{table*}[t]
{\normalsize
%
\begin{align} 
  & \text{\normalsize Thrust~\cite{Farhi:1977sg}$^{\decimal{footnote}}$:}
   \addtocounter{footnote}{1}
  & \tau & = \frac{1}{Q_p} \min_{\hat{t}} \sum_i \big(|\vec{p}_i|-
   |\vec{p}_i\cdot\hat{t}|\big)
  \,,
  & \bar\tau  = \frac{Q_p\,\tau}{Q} & = \frac{1}{Q}\sum_i p_i^\perp e^{-|\eta_i|}
   \,,\nn \\[3pt]
  & \text{\normalsize 2-jettiness~\cite{Stewart:2010tn}: } 
  & \tau_{2} &= \frac{1}{Q} \min_{\hat{t}} \sum_i \big(E_i - 
   |\vec{p}_i\cdot\hat{t}|\big)
   \,, 
  & \bar\tau_{2} =\tau_{2} &= \frac{1}{Q}\sum_i m_i^\perp e^{-|y_i|}
   \,. \nn \\[3pt]
  & \text{\normalsize Angularities~\cite{Berger:2003iw}$^{\decimal{footnote}}$: } 
     \addtocounter{footnote}{1}
  & \tau_{(a)} &=  \frac{1}{Q} \sum_i E_i \, (\sin\theta_i)^a
   (1-|\cos\theta_i|)^{1-a}
   \,,
  &\bar\tau_{(a)} =\tau_{(a)} 
   &= \frac{1}{Q}\sum_i p_i^\perp \frac{E_i}{|\vec{p}_i|} 
   \, e^{-|\eta_i|(1-a)}
   \,, \nn\\[3pt]
  & \text{\normalsize
    C-parameter~\cite{Parisi:1978eg,Donoghue:1979vi}$^{\decimal{footnote}}$:} 
  & C &= \frac{3}{2Q_p^2} \sum_{i,j} |\vec{p}_i||\vec{p}_j|\sin^2\theta_{ij}
   \,,
  &\bar C & = \frac{3}{Q}\sum_i \frac{p_i^\perp}{\cosh(\eta_i)} 
   \,, \nn\\[3pt]
  & \text{\normalsize Jet Masses \cite{Clavelli:1979md, Chandramohan:1980ry, Clavelli:1981yh}: } 
  & \rho_\pm  &=  \frac{1}{Q^2} \Big(\sum_{i\in \pm} p_i^\mu\Big)^2
   \,,
   & \bar\rho_\pm & = \frac{1}{Q}\sum_i m_i^\perp\theta(\pm y_i)\,e^{\mp y_i}
   \,, \nn
\end{align}}
\vspace{-0.5cm}
\caption{ \label{tab:classice}
Examples of event shapes $e$ with their original definitions, as well as
formulae $\bar e$ that are valid for $e \ll e_{\max}$.  See the text below \Eq{eq:Qdefinition} for a further description of the notation.
}
\end{table*}
\addtocounter{footnote}{-1} \addtocounter{footnote}{-1}
\footnotetext[\value{footnote}]{The original thrust variable is $T=1-\tau$.}
\addtocounter{footnote}{1} \footnotetext[\value{footnote}]{For $a = 1$,
  angularities reduce to jet broadening and hence are recoil sensitive.
  Throughout this paper we assume that $a < 1$ by an amount that allows 
  recoil effects to be neglected.}  \addtocounter{footnote}{1}
\footnotetext[\value{footnote}]{The shape parameter $H_2$ introduced in
  \cite{Fox:1978vu,Fox:1978vw} is equivalent to the C-parameter with the
  substitution $Q\to Q_p$.}
%

The remainder of this paper is organized as follows.  In
\Sec{sec:factorization}, we set the notation for event shapes and power
corrections.  In \Sec{subsec:transvelop}, we introduce the transverse velocity
operator whose matrix elements yield the leading power correction.  We explore
the consequences of universality classes in \Sec{sec:classes}, and use a
complete operator basis to derive approximate universality relations.  We
consider the effect of renormalization group evolution in \Sec{sec:running},
where we derive the anomalous dimension for $\Omega_1(r,\mu)$ and compare to the
hadronization models of Pythia~8 and Herwig++.  We conclude in
\Sec{sec:conclusions} with a discussion of the implications and extensions of
our results, in particular for the LHC.


\section{Power Corrections for Dijet Event Shapes}
\label{sec:factorization}
We begin by reviewing the notation for kinematics and event shapes in
\Sec{subsec:definitions}, various hadron mass schemes in \Sec{subsec:schemes},
and the basics of how power corrections impact event shapes in
\Sec{sec:observables}.  Readers familiar with these topics can skip to
\Sec{subsec:transvelop}, where we introduce the transverse velocity operator.

\subsection{Event Shapes with Transverse Velocity}
\label{subsec:definitions}

A dimensionless event shape $e$ is an observable defined on final-state particles
which can be used to describe the jet-like structure of an event.  To describe particle momenta, we use rapidity $y$, pseudo-rapidity
$\eta$, transverse momenta $p_\perp= |{\vec{p}_\perp}|$, and transverse mass $m^\perp$, where
$m_\perp = \sqrt{p_\perp^2+m^2}$ for a particle of mass $m$.  Defining
rapidities and transverse momenta relative to the $\hat z$ axis, a $4$-momentum
$p^\mu= (E,\vec p\,)$ can be written in two equivalent forms as
\begin{align}
  p^\mu &= 
  \big(m_\perp \cosh y, \vec p_\perp , m_\perp\sinh y\big)
   \\
  &= \Big( \sqrt{m^2+p_\perp^2\cosh^2\eta}, 
  \vec p_\perp , p_\perp\sinh\eta\Big) \,.\nn
\end{align}
In terms of the polar angle $\theta$ from the
$\hat z$ axis, $\eta=-\ln\tan(\theta/2)$. The standard velocity of a relativistic
particle is $v=|\vec p\,|/E$. In our analysis, an important role will be played by
the ``transverse velocity'' $r$ defined by
\begin{align}
  r = \frac{p_\perp}{m_\perp} = \frac{p_\perp}{\sqrt{p_\perp^2+m^2}} \,.
\end{align}
In general, the pseudo-rapidity and velocity of a particle can be expressed in
terms of $r$ and $y$:
\begin{align}\label{eq:eta-y-r}
\eta=\eta(r,y)  &= \ln\! \Bigg(\frac{\sqrt{r^2+\sinh^2 y}+\sinh y}{r}\Bigg)\,,
\\
v=v(r,y) & =\dfrac{\sqrt{r^2+\sinh^2y}}{\cosh y}\,.\nn
\end{align}
For massless particles, $r=v=1$ and $\eta=y$. 

\begin{table*}[t]
\begin{tabular}{l|ccccc}
$\boldsymbol{f_e(r,y)}$ & \hspace{0.8cm}$\tau$\hspace{0.8cm} 
  & \hspace{0.8cm}$\tau_{2}$ \hspace{0.8cm} 
  & \hspace{0.8cm}$\tau_{(a)}$ \hspace{0.8cm} 
  &  \hspace{0.8cm}$C$\hspace{0.8cm} 
  & \hspace{0.8cm}$\rho_{\pm}$\hspace{0.8cm}
 \\
\hline 
Original~~ 
  & $r\, e^{-|\eta|}$ 
  & $ e^{-|y|}$ 
  & $\dfrac{r}{v}\, e^{-|\eta|(1-a)}$ 
  & $\dfrac{3\, r}{\cosh\eta}$ 
  & $\theta(\pm y)\, e^{\mp y}$
  \\[3pt]
P-scheme~~ 
  & $r\, e^{-|\eta|}$ 
  & $r\, e^{-|\eta|}$ 
  & $r\, e^{-|\eta|(1-a)}$ 
  & $\dfrac{3\, r}{\cosh\eta}$ 
  & $r\,\theta(\pm\eta)\, e^{\mp\eta}$
  \\[3pt]
E-scheme~~ 
  & $\dfrac{r}{v}\, e^{-|\eta|}$ 
  & $\dfrac{r}{v}\, e^{-|\eta|}$ 
  & $\dfrac{r}{v}\, e^{-|\eta|(1-a)}$ 
  & $\dfrac{r}{v}\, \dfrac{3}{\cosh\eta}$ 
  & $\dfrac{r}{v}\, \theta(\pm\eta)\, e^{\mp\eta}$
  \\[3pt]
R-scheme~~ 
  & $r\, e^{-|y|}$ 
  & $r\, e^{-|y|}$ 
  & $r\, e^{-|y|(1-a)}$ 
  & $\dfrac{3\, r}{\cosh y}$
  & $r\, \theta(\pm y)\, e^{\mp y}$
  \\[3pt]
J-scheme~~ 
  & $ e^{-|y|}$ 
  & $ e^{-|y|}$ 
  & $e^{-|y|(1-a)}$ 
  & $\dfrac{3}{\cosh y}$ 
  & $\theta(\pm y)\, e^{\mp y}$
\end{tabular}
\caption{\label{tab:f-functions}
  The functions $f_e(r,y)$ of ``transverse velocity'' $r$ and rapidity $y$ for 
  various dijet event shapes using several different schemes for treating 
  hadron mass effects. Here $\eta=\eta(r,y)$ and $v=v(r,y)$ are given in 
  \eq{eta-y-r}.}
\end{table*}

Our focus will be on dijet event shapes in $e^+e^-$ collisions, which have the property that $e\to 0$ implies back-to-back pencil-like
jets (and $e=0$ for the lowest order partonic configuration $e^+e^-\to q\bar
q$). To maintain simplicity, we will not include recoil sensitive observables
such as broadening \cite{Catani:1992jc} in our analysis.\footnote{For a recoil sensitive observable,
  the axis used to compute $e$ (typically the thrust axis $\hat t$) differs from the axis which
  minimizes $e$ by an amount that can have an ${\cal O}(1)$ effect on the value
  of $e$.}  The event shapes we consider are also bounded as $0 \le e \le
e_{\max}$, where the maximum value $e_{\max}$ depends on the observable in question but is typically $\mathcal{O}(1)$.
For the $e\ll e_{\max}$ limit, one can sometimes simplify the expression defining
$e$ by neglecting corrections of ${\cal O}(e^2)$. We will distinguish
expressions for event shapes that are valid under this approximation by adding a
bar, $\bar e$.

Various examples of dijet event shapes are shown in \tab{classice}, including
their original definitions $e$ and expressions $\bar e$ valid when $e\ll e_{\max}$.  For angularities and 2-jettiness, there are no
simplifications in the dijet limit, so $\bar e=e$ without higher order terms.  The
notation for various items in the table require explanation.
The normalization factors are defined by
\begin{align}
\label{eq:Qdefinition}
 Q_p&=\sum_i |\vec p_i|\,,
 & Q &=\sum_i E_i \,.
\end{align}
The unit vector $\hat t$ obtained in the minimization defining thrust $\tau$ is
referred to as the thrust axis, and we have taken it to be aligned with $\hat z$ to
define rapidities. The angle between particles $i$
and $j$ is denoted by $\theta_{ij}$, and the angle between particle $i$ and $\hat t$ is denoted by $\theta_i$.  Finally,
the $\pm$ labels refer to each of the two hemispheres defined by the plane normal to
the thrust axis. For later convenience, we have written all $\bar e$ formulae in
terms of $p_\perp$, $m_\perp$, $y$, and/or $\eta$.  Note that we can always
replace $Q_p\to Q$ in the overall normalization for $\bar e$, since the
correction in doing so is beyond the order to which we are working.

For a dijet event shape $e$, the largest part of the cross section comes from $e\ll
e_{\rm max}$, as do the most important hadronization corrections which are the focus of this paper.
The simplicity of the dijet limit makes it possible to derive factorization
theorems\footnote{The simplest examples of dijet factorization rely on being
  able to write the dijet event shape as a sum of contributions from energetic
  collinear particles in the $\pm$ hemispheres ($n$,\,$\bar n$), soft perturbative
  particles ($s$), and soft nonperturbative particles ($\Lambda$), via $\bar e =
  e_{n} + e_{\bar n} + e_s + e_\Lambda$. See for example \Ref{Bauer:2008dt}.} for
these cross sections which facilitate calculations of higher order perturbative
corrections $\propto \alpha_s^j (\ln^ke)/e$, as well as defining nonperturbative
corrections in terms of field-theoretic matrix elements.  For our purposes, the
relevant point is that at leading order in the nonperturbative corrections, we
can split $\bar e$ into perturbative ($e_{\rm p}$) and non-perturbative
($e_\Lambda$) contributions\footnote{The heavy jet mass event shape
  $\rho_H=\max\{\rho_+,\rho_-\}$ does not admit such a decomposition, and
  correspondingly its factorization formula is a bit more complicated. The
  light jet mass $\rho_L=\min\{\rho_+,\rho_-\}$ is not a true dijet event shape
  since $\rho_L\to 0$ does not imply a dijet configuration.}
\begin{align} \label{eq:splite}
  \bar e \,=\, e_{\rm p} \,+\, e_{\Lambda} \,.
\end{align}
The $e_{\rm p}$ term is generated by particles with momenta $p^\mu\gg \Lambda_{\rm
  QCD}$ and here we can neglect corrections from hadron masses $m_H$ up to second order in
the $m_H/Q$ expansion.  The $e_\Lambda$ term involves corrections from soft
particles with momenta $p^\mu \sim \Lambda_{\rm QCD}$, and from
\tab{classice}, we see that $e_\Lambda \sim \Lambda/Q$ where $\Lambda\sim
m_H\sim \Lambda_{\rm QCD}$. Thus simple power counting dictates that to
determine $e_{\Lambda}$, we cannot neglect hadron mass effects in the
definition of $\bar{e}$.

For our analysis, we will find it convenient to characterize each event shape by
a function $f_e(r,y)$ of the transverse velocity $r$ and rapidity $y$, which we define
from the $e\ll e_{\max}$ limit via
\begin{align}\label{eq:f-definition}
\bar e = \frac{1}{Q}\sum_{i} m^\perp_i f_e(r_i, y_i)\,.
\end{align}
The various examples in \tab{classice} have the following $f_e(r,y)$ functions:
\begin{align}
 f_\tau(r,y) &= \sqrt{r^2+\sinh^2(y)}-\sinh|y|
   \,, \\
 f_{\tau_{2}}(r,y) &=  e^{-|y|}
  \,,\nn \\
 f_{\tau_{(a)}}(r,y) &= r^a\,\frac{\partial\eta(r,y)}{\partial y}
   \bigg(\sqrt{r^2+ \sinh^2y}-\sinh|y|\bigg)^{1-a}
   ,\nn\\
 f_C(r,y) &= \dfrac{3\,r^2}{\sqrt{r^2+\sinh^2y}}
   \,,\nn\\
 f_{\rho_\pm}(r,y) &= \theta(\pm y) \, e^{\mp y}
   \,,\nn
\end{align}
where
\begin{align}
\label{eq:vrelations}
\frac{E}{|\vec{p}\,|} & = \frac{\cosh y }{\sqrt{r^2+\sinh^2 y}} =
\frac{\partial \eta(r,y)}{\partial y} = \frac{1}{v}\,.
\end{align}
The translation from the notation of \Ref{Salam:2001bd} (denoted with
superscripts ${\rm SW}$) to our notation is $f_e^{\rm SW}(y,m^2/p_\perp^2) =
f_e(r,y)/r$.

\subsection{Hadron Mass Schemes}
\label{subsec:schemes}
%
%
In experimental analyses, different ``schemes'' are often adopted for the
treatment of hadron masses depending on the available information. In the
context of event shapes, a detailed discussion of these schemes is given in
\Ref{Salam:2001bd}.  These schemes correspond to different choices for
$f_e(r,y)$ that yield the same value for the perturbative contribution $e_{\rm p}$ but
potentially different values for the nonperturbative contribution $e_\Lambda$.
The schemes considered in this paper are summarized in \tab{f-functions}.

In the ``P-scheme'', one makes measurements with only $3$-momentum information.
One performs the substitution $E_i \to |\vec{p}_i|$ in the formula for event
shapes, and correspondingly $m_\perp \to r\, m_\perp$ in \eq{f-definition}.  To
satisfy infrared safety, the original $f_e(r,y)$ must tend to a constant value
(possibly zero) in the $r\to 0$ limit, and this implies that the P-scheme event
shapes will always vanish linearly with $r$.  The P-scheme replacement affects
the jet masses, angularities, and 2-jettiness which become
\begin{align}
\tau_{2}^P & 
   = \frac{1}{Q_p} \sum_i p_i^\perp e^{-|\eta_i|}=  \tau 
   \,, \\
 \tau_{(a)}^P &= \frac{1}{Q_p}\sum_i |\vec{p}_i| \, (\sin\theta_i)^a
   (1-|\cos\theta_i|)^{1-a} \,,
   \nn\\
 \rho_\pm^P &= \frac{2}{Q_p^2}\sum_{(i,j)\in \pm} |\vec{p}_i||\vec{p}_j|\sin^2
 \frac{\theta_{ij}}{2} 
  \,. \nn
\end{align}

Another scheme discussed in \Ref{Salam:2001bd} is the ``E-scheme'' (see
\Refs{Korchemsky:1994is, Korchemsky:1999kt, Sveshnikov:1995vi,
  Tkachov:1999py} for earlier discussions), where only measurements of energies
and angles are used to construct observables.  Compared to the P-scheme, one makes the substitution
\begin{align}\label{eq:Escheme}
\vec{p}_i \to \frac{E_i}{|\vec{p}_i|}\,\vec{p}_i \,,
\end{align}
in the formula for event shapes.  This modifies all examples in \tab{classice}
except for angularities: 
\begin{align}
\tau^E = \tau_2^E &= \frac{1}{Q}\sum_i E_i(1-|\cos\theta_i|)\,,
\\
\rho^E_\pm &= \frac{2}{Q^2}\sum_{(i,j)\in \pm} E_i E_j
  \sin^2 \frac{\theta_{ij}}{2}\,,
  \nn\\
C^E &= \frac{3}{2Q^2}\sum_{i,j} E_iE_j\sin^2\theta_{ij}\,.\nn
\end{align}
Note that thrust and 2-jettiness are identical in the \mbox{E-scheme},
$\tau_2^E=\tau^E$.  In the P-scheme, the event shapes 
are all linear in $r$, and hence the corresponding E-scheme results for
$f_e(r,y)$ are simply obtained by multiplying by $E/|\vec{p}\,|=1/v$.  Note that
the E-scheme and P-scheme are defined in terms of pseudo-rapidity $\eta$
(equivalently, the polar angle $\theta$).


In order to consider a wider range of observables, we will introduce two new
schemes.  In the ``R-scheme'' (rapidity scheme), we take event shapes defined in
the \mbox{P-scheme} and make the replacement $\eta\to y$.  To define R-scheme event
shapes where $\bar e\ne e$, we carry out this replacement for $\bar e$, and then
define $e^R=\bar e^R$.  The \mbox{``J-scheme''} is the closest to the jet mass observables,
and is defined by taking the R-scheme result and setting $r=1$,
$f_{e^J}(r,y)=f_{e^R}(1,y)$.

We emphasize that the naming of schemes discussed here is set simply by
convention. For understanding power corrections one only needs to know the
functional form of $f_e(r,y)$.

\subsection{Effect of Power Corrections on Cross Sections}
\label{sec:observables}
For recoil-less dijet event shapes that satisfy \eq{splite},
the perturbative/nonperturbative factorization of the differential
distribution in the $e\to 0$ limit implies
\begin{equation}\label{eq:fact-thmn}
 \frac{\df\sigma}{\df e}=\int \!\df \ell \:
   \frac{\df\hat\sigma}{\df e}\bigg(e-\frac{\ell}{Q}\bigg)\,F_e(\ell) 
  \, \big[ 1 +{\cal O}(e) \big] 
  \,.
\end{equation}
Here $\df\hat{\sigma}/\df e$ is the most singular perturbative cross section to
all orders in $\alpha_s$, and contains the full leading power perturbative soft
function.  $F_e$ is the shape function that depends on the specific event shape
one is interested in. It contains nonperturbative power corrections (and, as we
will see, perturbative corrections). If \mbox{$Qe\sim \Lambda_{\rm QCD}$}, then
the entire function $F_e(\ell)$ has an important impact on the cross section and
in practice one models it with a few coefficients which can be fit to
data~\cite{Korchemsky:1999kt,Korchemsky:2000kp}, or uses a complete basis which
can be systematically improved~\cite{Ligeti:2008ac}.

For $Q e\gg \Lambda_{\rm QCD}$, the function $F_e(\ell)$ can be expanded for $\ell\gg \Lambda_{\rm QCD}$ in terms of
nonperturbative matrix elements of operators. The first terms are
\begin{align}\label{eq:shape-function}
F_e(\ell) &= \delta(\ell)  -\delta^\prime(\ell) \,\Omega_1^e 
  +{\cal O}\Big(\frac{\alpha_s\Lambda_{\rm QCD}}{\ell^2}\Big) 
  + {\cal O}\Big(\frac{\Lambda_{\rm QCD}^2}{\ell^3}\Big) , 
\end{align}
where $\Omega_1^e(\mu)$ is a dimension-$1$ nonperturbative matrix element (defined
here in the $\overline{\rm MS}$ scheme) that encodes the power corrections we
wish to study.  It is defined by
\begin{align}\label{eq:Omega-definition}
\Omega^e_1 &= \langle\,0\,|\, \overline{Y}_{\bar n}^\dagger Y_n^\dagger
    (Q\hat e) Y_n \overline{Y}_{\bar n}\,|\,0\,\rangle\,,
\end{align}
where $(Q\hat e)$ is a $Q$-independent field-theoretic operator that measures the
combination $Q\bar e$, and $Y$ ($\overline{Y}$) are Wilson lines with gluon
fields in the fundamental (anti-fundamental) color representation along the
directions specified by $n=(1,\hat t)$ and \mbox{$\bar n=(1,-\hat t)$}. For example,
\begin{equation}\label{eq:wilson-lie}
Y_n = {\rm P} \exp \Bigg[ ig\!\int_0^\infty\! \df s\,n\cdot A(ns)\Bigg]\,,
\end{equation}
where P stands for path-ordering and $A^\mu$ is the gluon field.  The fact that the
dimension-$1$ matrix element $\Omega_1^e$ is nonperturbative is easy to
understand on dimensional grounds since the only scale for this QCD vacuum
matrix element is $\Lambda_{\rm QCD}$.  In \eq{shape-function} the ${\cal
  O}(\alpha_s\Lambda_{\rm QCD}/\ell^2)$ term involves perturbative corrections to
the leading power correction which will be discussed in \Sec{sec:ope}.

Plugging \Eq{eq:shape-function} into \Eq{eq:fact-thmn} for the event
shape distribution one finds
\begin{equation} \label{eq:distnshift}
\frac{\df\sigma}{\df e} = \frac{\df \hat\sigma}{\df e} 
  - \frac{\Omega_1^e}{Q} 
 \frac{\df}{\df e} \frac{\df \hat\sigma}{\df e} +\ldots\,,
\end{equation}
where the ellipsis denote higher order terms in $\alpha_s$ and $\Lambda_{\rm
  QCD}/\ell$. \Eq{eq:distnshift} corresponds to a shift $e\to
e\,-\,\Omega_1^e/Q$ to first order in $1/Q$, and reproduces the known shift found in the
dispersive approach \cite{Dokshitzer:1995zt,Dokshitzer:1995qm,Dokshitzer:1998pt}. Thus, the dominant effect of power corrections
(hadronization) on dijet event shapes for $Q e\gg \Lambda_{\rm QCD}$ is simply a shift in the distribution.

Following \Ref{Abbate:2010xh} we note that as long as $\df
\hat{\sigma}/\df e$ tends to zero in the far tail region, one can derive an operator product expansion for the first moment of
the distribution as well.  Defining the full and perturbative moments as
\begin{equation}
\label{eq:defpertmoment}
\langle e \rangle  \equiv  \int \!\df e\: e\,\frac{1}{\sigma}\frac{\df \sigma}{\df e}\,, \qquad
\langle e \rangle_\pert  \equiv  \int \!\df e\: e\,\frac{1}{\hat\sigma}\frac{\df \hat \sigma}{\df e}\,,
\end{equation}
one can use the factorization in \eq{fact-thmn} and the expansion of the shape function in \eq{shape-function} to show
\begin{equation}
\label{eq:momshift}
\langle e \rangle  =  \langle e \rangle_\pert  +
\frac{\Omega_1^e}{Q} + \ldots \,,
\end{equation}
where again the ellipsis denotes higher order terms.  For the event shapes in
\tab{classice}, all the event shapes except for $C$ tend to zero in the far tail region,
so the leading power correction generates a $Q$-dependent shift of their
first moment. We will use this feature to extract $\Omega_1^e$
from the Pythia~8 and Herwig++  hadronization models in \Sec{sec:MC}.
See Ref.~\cite{Korchemsky:2000kp} for a discussion of the modification necessary for the C-parameter moment.

\section{The Transverse Velocity Operator}
\label{subsec:transvelop}
The goal of this paper is to study the effect of hadron masses on the power correction in \Eq{eq:Omega-definition}.
In order to formulate the operator $(Q\hat e)$ appearing in the definition of
$\Omega_1^e$, we will follow the energy-momentum tensor approach of
\Refs{Lee:2006nr,Bauer:2008dt}, and generalize it so that we can treat the dependence on the
transverse velocity $r$.  In particular, the event shape $e$ will be written as the
eigenvalue of an operator acting on the final state which includes hadron
mass effects.   In \Sec{subsec:boost}, we consider the crucial role of boost invariance for identifying universality classes.

\subsection{Comparison to Transverse Energy Flow}
\label{subsec:trans-en-flow}
To set up our analysis, it will be convenient to first review definitions from the
literature that do not account for hadron mass effects in the event shapes.
These correspond to setting $r=1$ for the event shapes defined by \Eq{eq:f-definition}. The transverse energy
flow operator $\hat {\cal E}_T(\eta)$ is defined by~\cite{Lee:2006nr}
\begin{align} \label{eq:defineET}
  \hat {\cal E}_T(\eta) | X \rangle 
  = \sum_{i\in X} p_i^\perp\, \delta(\eta-\eta_i) |X\rangle  \,.
\end{align}
(Here and below we suppress the dependence on the thrust axis $\hat t$.) 
The operator $\hat {\cal E}_T(\eta)$ is
related to the energy-momentum tensor $T^{\mu \nu}(t, \vec{x})$
by~\cite{Sveshnikov:1995vi,Korchemsky:1997sy,Belitsky:2001ij,Bauer:2008dt}
\begin{align} \label{eq:ETwithT}
  \hat {\cal E}_T(\eta) 
   = \frac{1}{\cosh^3\!\eta} \int_0^{2\pi}\!\!\!\!\! \df\phi
   \lim_{R\to\infty} R^2\! \int_0^\infty \!\!\!\! \df t\: 
   \hat n_i\, T^{0i}(t,R\,\hat n)
   ,
\end{align}
where $\hat n$ is a unit vector pointing in the $(\theta,\phi)$ direction for
the $\theta$ corresponding to $\eta$.\footnote{In the proof of \Ref{Bauer:2008dt}
in which \Eq{eq:ETwithT} reproduces \Eq{eq:defineET} for scalars and
fermions, one assumes that all particles are massless. If the energy momentum
tensor is considered for massive fields, then \Eq{eq:ETwithT} yields \Eq{eq:defineET},
but $\eta$ must be identified with pseudo-rapidity and the factor $p_i^\perp\to
m_i^\perp$.}  Using $\hat {\cal E}_T(\eta)$, we can define an operator for the event shape $\bar e$ by~\cite{Lee:2006nr}
\begin{align}
 \hat e_0 | X\rangle 
  \equiv \frac{1}{Q} \int\! \df\eta\, f_e(\eta)\, \hat {\cal E}_T(\eta) |X\rangle \,,  \label{eq:e0}
\end{align}
where in the notation defined in \eq{f-definition}, this
\mbox{$f_e(\eta)=f_e(r=1,y=\eta)$}.  For massless particles with \mbox{$r=1$}, this operator satisfies $\hat e_0 |X\rangle =
\bar{e}(X)|X\rangle$, so the measurement operator in \Eq{eq:Omega-definition} is given by
$\widehat {\cal M}_e = Q \hat e_0$.

To generalize this operator formalism to include hadron masses, we require a
transverse momentum flow operator that is more differential, namely a
$\Eop(r,y)$ which can pick out states that have a particular transverse velocity
$r$.  We will refer to this simply as the ``transverse velocity operator'', and
define it by its action on a state
\begin{equation}\label{eq:trans-mom-flow-op}
 \Eop(r, y)\! \ket{X} 
  = \sum_{i \in X} m_i^\perp\delta(r-r_i)\, \delta(y - y_i) \ket{X}\,.
\end{equation}
Note that here we use rapidity $y$ rather than pseudo-rapidity $\eta$.  In \App{sec:derivation} we show that $\Eop(r,y)$ can be defined in
terms of the energy-momentum tensor as
\begin{align}\label{eq:Eop-definition}
\Eop(r,y) & 
  =
 \frac{r\: {\rm sech}^4y}{\sqrt{r^2 \!+\! {\rm sinh}^2y}}
    \lim_{R\to\infty}\! R^3 \!\! \int_0^{2\pi}\!\!\!\!\! \df \phi\: 
    \hat{n}_i\, T^{0i}(R,R\, v\,\hat{n})\,,
\end{align}
where $v=v(r,y)$ and $\eta=\eta(r,y)$ are given in \eq{eta-y-r}.
The unit vector $\hat{n}$ again points in the $(\theta,\phi)$ direction and hence
depends on $y$ and $r$ through its dependence on \mbox{$\eta =
  -\ln\tan(\theta/2)$}.

\begin{figure}
\begin{center}
\includegraphics[width=1.0\columnwidth]{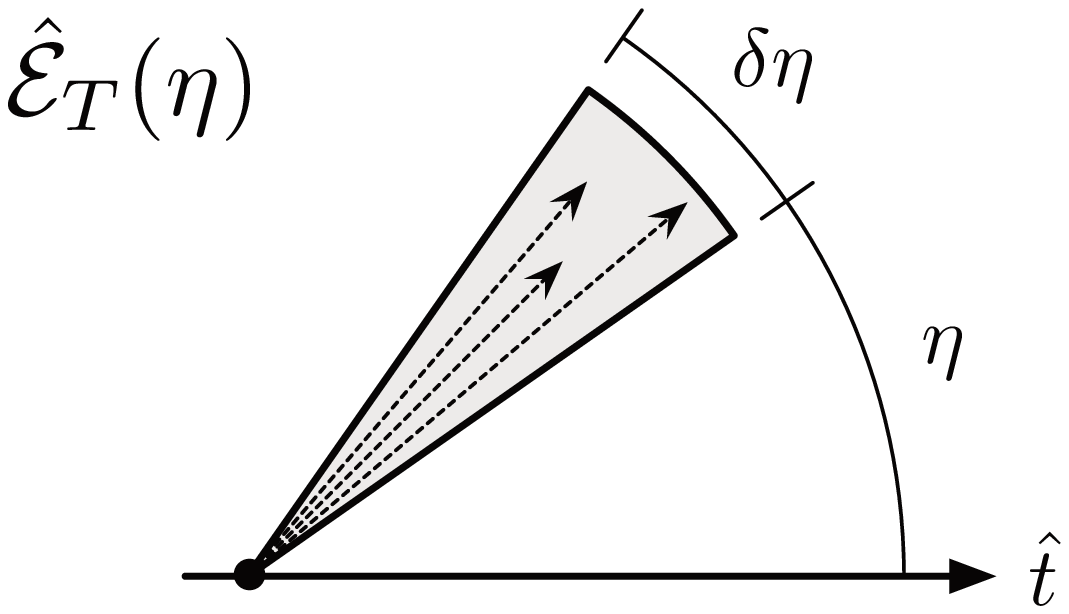} \\
\includegraphics[width=1.0\columnwidth]{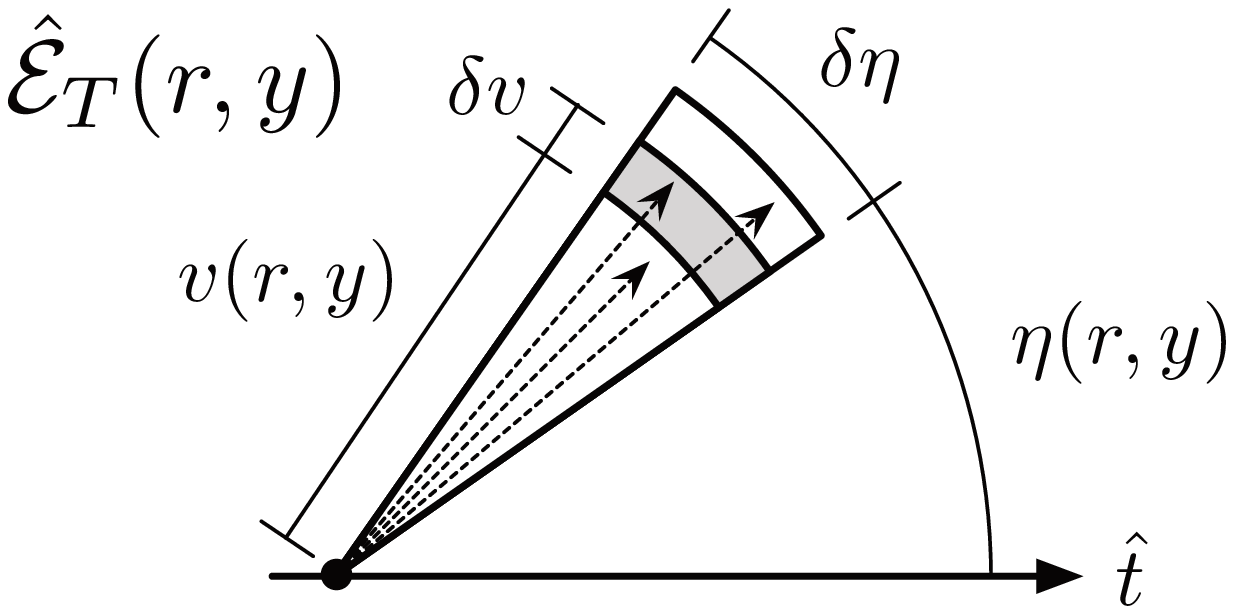}
\end{center}
\caption{The energy flow operator (top) compared to the transverse velocity operator (bottom).  Measurements are made with
respect to the thrust axis $\hat{t}$.  The arrows correspond to particles with lengths given by the particle velocities. 
Shading indicates which particles are measured by the operator.  Note that  the velocity $v(r,y)$ and pseudo-rapidity $\eta(r,y)$
are functions of the transverse velocity $r$ and rapidity $y$.}
\label{fig:transvelocityop}
\end{figure}

The physical picture for the distinction between $\hat
{\cal E}_T(\eta)$ and $\Eop(r,y)$ is shown in \Fig{fig:transvelocityop}.
The energy flow operator
$\hat {\cal E}_T(\eta)$ involves an expanding sphere of
radius $R$ integrated over all time, and measures the total transverse momentum
for rapidities in an infinitesimal interval $\delta\eta$ about $\eta$.  The
transverse velocity operator $\Eop(r,y)$ involves a spheroid that expands in
both space and time with a finite velocity $v$, and it measures the total
transverse mass for particles in an infinitesimal interval in both $\eta$ and the
velocity $v$ (or equivalently an infinitesimal interval in $y$ and $r$).  
Using $\Eop(r,y)$, the value of an event shape $\bar e$ for a state $|X\rangle$
with massive or massless particles is given by \mbox{$\hat e |X\rangle =
\bar{e}(X)|X\rangle$} where the operator
\begin{align} \label{eq:e}
\hat{e} & \equiv \dfrac{1}{Q}\int_{-\infty}^{+\infty}\!\!\! \df y\,
  \int_0^1\!\! \df r \: f_e(r, y)\, \Eop(r, y) \,,
\end{align}
involves $f_e(r,y)$ defined in \eq{f-definition}. This is the desired
generalization of \eq{e0} that will allow us to treat the effect of hadron
masses on event shape power corrections.  The result in \eq{e} completes the
matrix element definition of $\Omega_1^e$ given in \eq{Omega-definition}.


\subsection{Boost Invariance}
\label{subsec:boost}
Both $\hat {\cal E}_T(\eta)$ and $\Eop(r,y)$ have nice transformation properties
under longitudinal boosts.  These arguments were first given in \Ref{Lee:2006nr}
in the context of $\hat {\cal E}_T(\eta)$ to prove universality of power
corrections for massless particles, and we will extend the logic for $\Eop(r,y)$
to develop the notion of universality classes which account for hadron masses in
the next section.

For the case of massless particles in \Ref{Lee:2006nr}, $\eta=y$ and
it was shown that under a boost of rapidity $y^\prime$ along the
thrust axis:
\be
U(y') \hat {\cal E}_T(y) U(y')^\dagger = \hat {\cal E}_T(y+y').
\ee
Due to the invariance of the vacuum $|0\rangle$ and Wilson lines $Y_n$ and
$Y_{\bar n}$ under this boost, we can choose $y^\prime = -y$ so
\begin{align}
\langle \,0 \,|\, \overline{Y}_{\bar n}^\dagger Y_n^\dagger
    {\cal E}_T(y) Y_n \overline{Y}_{\bar n} \,| \,0 \,\rangle 
  = \langle \,0 \,|\, \overline{Y}_{\bar n}^\dagger Y_n^\dagger
    {\cal E}_T(0) Y_n \overline{Y}_{\bar n} \,| \,0 \,\rangle.
\end{align}
Using \Eq{eq:e0}, again with $\eta=y$, the power correction in
\Eq{eq:Omega-definition} simplifies to
\be
\label{eq:masslessuniversality}
\left[\int_{-\infty}^{+\infty}  \! \df\eta\, f_e(\eta)\right] \langle \,0 \,|\, \overline{Y}_{\bar n}^\dagger Y_n^\dagger
    {\cal E}_T(0) Y_n \overline{Y}_{\bar n} \,| \,0 \,\rangle \,.
\ee
Thus as argued in \Ref{Lee:2006nr}, the power correction is universal if one
neglects hadron mass effects, since it depends on a common nonperturbative
matrix element times a calculable $\eta$ integral specific to an event shape.
This result agrees with the dispersive approach~\cite{Dokshitzer:1995zt,
Dokshitzer:1995qm,Dokshitzer:1998pt}, and also explains
why it only applies for the first power correction in the expansion of
\eq{shape-function}.

We can apply the same logic to our transverse velocity operator in
\Eq{eq:trans-mom-flow-op}, now accounting for the effect of hadron masses.
Under a boost of rapidity $y^\prime$,
\begin{equation}\label{eq:boost-transformation}
U(y^\prime)\,\Eop(r, y)\,U(y^\prime)^\dagger= \Eop(r, y  + y^\prime)\,.
\end{equation}
Choosing $y^\prime = -y$, we find that the leading power corrections to dijet
event shapes are all described by the nonperturbative matrix element
\be
\label{eq:universalitya}
\Omega_1(r) \equiv \langle \,0 \,|\,  
   \overline{Y}_{\bar n}^\dagger Y_n^\dagger 
    \Eop(r, 0) Y_n \overline{Y}_{\bar n}
   \,| \,0\,\rangle
\ee
which depends only on $r$ and is independent of the event shape $e$. Using
\Eq{eq:e}, the power correction $\Omega_1^e$ simplifies to
\begin{align} \label{eq:O1univ}
  \Omega_1^e 
  &= \int_0^1\!\! \df r
   \bigg[\!\int_{-\infty}^{+\infty}\!\!\!\!\! \df y\, f_e(r,y)\! \bigg] 
   \Omega_1(r) \,.
\end{align}
We will consider the implications of \Eqs{eq:universalitya}{eq:O1univ}
for universality in the next section.  Note that there is no limit where \Eq{eq:O1univ}
approaches the massless approximation in \Eq{eq:masslessuniversality}.

\section{Universality for Event Shapes}
\label{sec:classes}
\begin{table*}[t!]
\begin{tabular}{l|ccccc}
 $\boldsymbol{c_e}$ 
  & \hspace{0.8cm}$\tau$\hspace{0.8cm} 
  & \hspace{0.8cm}$\tau_{2}$ \hspace{0.8cm} 
  & $\hspace{0.8cm}\tau_{(a)}\hspace{0.8cm}$ 
  & \hspace{0.8cm}$C$\hspace{0.8cm} 
  & \hspace*{0.8cm}$\rho_{\pm}\hspace*{0.8cm}$\\
\hline 
Common~~ 
  & $2$
  & $2$ 
  & $\dfrac{2}{1-a}$ 
  & $3\pi$ 
  & $1$
  \\[3pt]
\end{tabular}
\caption{\label{tab:cfunctions}
Expression for the $c_e$ coefficients for various dijet event shapes from \Tab{tab:f-functions}. Since $c_e$ are defined using
$f_e(1,y)$, they have the same value in each class. }
\end{table*}
\begin{table*}[t!]
\begin{tabular}{l|ccccc}
 $\boldsymbol{g_e(r)}$ & \hspace{0.8cm}$\tau$\hspace{0.8cm} 
  & \hspace{0.8cm}$\tau_{2}$ \hspace{0.8cm} 
  & $\hspace{0.8cm}\tau_{(a)}\hspace{0.8cm}$ 
  & \hspace{0.8cm}$C$\hspace{0.8cm} 
  & \hspace*{0.8cm}$\rho_{\pm}\hspace*{0.8cm}$
  \\
\hline 
Original~~ 
  & $g_\tau(r)$
  & $1$ 
  & $r$ 
  & $\frac{2r^2}{\pi}K(1-r^2)$ 
  & $1$
  \\[3pt]
P-scheme~~ 
  & $g_\tau(r)$ 
  & $g_\tau(r)$ 
  & $g_{\tau_a}(r)$ 
  & $\frac{2r^2}{\pi}K(1-r^2)$ 
  & $g_\tau(r)$
  \\[3pt]
E-scheme~~ 
  & $r$ 
  & $r$
  & $r$
  & $r$
  & $r$
  \\[3pt]
R-scheme~~ 
  & $r$ 
  & $r$
  & $r$
  & $r$
  & $r$
  \\[3pt]
J-scheme~~ 
  & $1$ 
  & $1$ 
  & $1$ 
  & $1$ 
  & $1$
\end{tabular}
\caption{\label{tab:g-functions}
  The functions $g_e(r)$ of transverse velocity $r$ for  various dijet event shapes from \Tab{tab:f-functions}.
Event shapes with the same $g(r)$ belong to the same universality class.}
\end{table*}
\subsection{Universality Classes Defined by $\boldsymbol{g(r)}$}
\label{sec:gclasses}
The boost invariance logic of the previous section shows that for the power correction $\Omega_1^e$, we can factor out
the rapidity dependence from the nonperturbative matrix element. We define the
integral appearing in \eq{O1univ} as
\begin{align}
  \int_{-\infty}^{+\infty}\!\!\! \df y\ f_e(r,y)\, \equiv\, c_e\: g_e(r) \,,
\end{align} 
where $g_e(1)=1$ and 
\begin{align} \label{eq:defncg}
&c_e = \int_{-\infty}^{+\infty}\!\!\!\!\! \df y \,f_e(1, y)\,,
&g_e(r) & = \frac{1}{c_e}\int_{-\infty}^{+\infty}\!\!\!\!\! \df y \,f_e(r, y)\,.
\end{align}
(For the special case of event shapes that vanish for massless partons, $\int
\df y f_e(1,y)=0$, we define $c_e=1$ and let $g_e(1)=0$.)
Thus, the leading power correction for an event
shape $e$ can be written as
\begin{align}\label{eq:universality}
&\Omega^e_1 = c_e \,\Omega^{g_e}_1, \qquad \Omega^{g_e}_1 \equiv \int_0^1\!\! \df r \,g_e(r)\,\Oop\,,
\end{align}
where $\Omega_1(r)$ is given in \eq{universalitya}.

Using the fact that for massless hadrons $f_e(\eta) = f_e(r=1, y=\eta)$ in
\Eq{eq:masslessuniversality}, we see that the coefficients $c_e$ are precisely
the classic universality prefactors that one derives neglecting the hadron mass
dependence of event shapes~\cite{Korchemsky:1994is, Dokshitzer:1995zt,
  Dokshitzer:1995qm, Dokshitzer:1997ew, Korchemsky:1999kt, Belitsky:2001ij,
  Berger:2003pk, Berger:2004xf, Lee:2006nr}.  The function $g_e(r)$ then encodes
the effect of hadron masses through the nonperturbative parameter
$\Omega^{g_e}_1$.

The key result from \Eq{eq:universality} is that each unique function $g_e(r)$
defines a universality class for dijet event shapes.  In particular,  
for two different event shape variables $a$ and $b$, if $g_a(r)=g_b(r)$ then
$\Omega^{g_a}_1 = \Omega^{g_b}_1$ (equivalently $\Omega^a_1 / c_a = \Omega^b_1 / c_b$), so their power corrections agree up to the
calculable constants $c_{a}$ and $c_{b}$.  We say that two such event shapes belong to the same universality class,
and we will often refer to $\Omega^{g}_1$ as the universal power correction defined by $g(r)$.

Recall from \tab{classice} that the $f_e(r,y)$ functions in general depend on
the measurement scheme used for treating hadron mass effects (E-scheme,
P-scheme, etc). The functions $g_e(r)$ will also in general vary with the
measurement scheme. However, the $c_e$ coefficients are independent of the scheme
for treating hadron masses since they are defined with $r=1$ corresponding to
the massless limit, and \mbox{$f_e(r=1,y)$} is the same in all schemes.  Our
classification of universality classes with common $g(r)$'s is the same as the
identification of event shapes that have the same hadron mass effects made in
\Ref{Salam:2001bd}, and we will elaborate on the precise notational
relationship in \Sec{sec:classic_event_shapes} below.

A summary of coefficients $c_e$, functions $g_e(r)$, and universality classes
is given in Tables~\ref{tab:cfunctions}, \ref{tab:g-functions},
and~\ref{tab:classes}, and will be discussed in detail in 
\Secs{sec:toy-event-shapes}{sec:classic_event_shapes} below.
Since many of the standard event shapes have different $g(r)$ functions, their
power corrections are not related by universality. We will take up the question
of numerically approximate relations between power corrections in different
universality classes in \Sec{sec:basis}. Finally in
\Sec{sec:renormalon} we consider the impact of changing the
renormalization scheme defining $\Omega_1(r)$ from $\overline{\rm MS}$ to a
renormalon-free scheme.

\subsection{Generalized Angularities}
\label{sec:toy-event-shapes}
In order to see how universality works in practice, it is instructive to
consider a family of event shapes which are a simple generalization of
angularities, and are labeled by two numbers $n \geq 0$ and $a < 1$:
\begin{align} \label{eq:tna}
  \ang{n,a} &\equiv  \sum_i m_i^\perp r_i^n e^{-|y_i| (1-a)}\,.
\end{align}
This corresponds to extending the R-scheme definition of angularities by
incorporating a positive power $n$ of $r=p_\perp/m_\perp$, and this $n$
dependence allows the event shape to directly probe hadron mass effects. In
terms of \Eq{eq:f-definition} one trivially finds
\begin{equation}
f_{n,a}(r,y) = r^n e^{-|y|(1-a)}\,.
\end{equation}
These generalized angularities all have  the same value of $c_e$,
\begin{equation}
\label{eq:cadef}
c_{n,a} = \int_{-\infty}^{+\infty}\!\!\!\!
  \df y \, e^{-|y|(1-a)} = \frac{2}{1-a} \equiv c_a \,,
\end{equation}
which from \Tab{tab:cfunctions} is also the same as for classic angularities
$\tau_{(a)}$. (Again $a<1$ here.)
Computing the function encoding the mass dependence we have
\begin{align}
g_{n,a}(r) = \frac{(1-a)}{2}\int \df y \,  r^n e^{-|y|(1-a)} = r^n \,.
\end{align}
Thus the $\ang{n,a}$ event shapes belong to universality classes labeled by $n$,
and represented by the functions
\begin{equation}
g_n(r) = r^n \,. 
\end{equation}
Each value of $n$ defines a different universality class, so in general there
are infinitely many different event shape universality classes.  For the $r^n$ class,
the universal power correction from \Eq{eq:universality} is
\be
\label{eq:universalrn}
\Omega_1^{n} \,\equiv\, \int_0^1 \! \df r \, r^n \, \Omega_1(r).
\ee 

This procedure, of generalizing an event shape to obtain different sensitivity
to hadron mass effects by multiplying by $r^n$, can just as easily be applied to
event shapes other than angularities.  By choosing large values of $n$, we
preferentially select out particles whose transverse momentum is large compared
to its mass.  This might be useful in an experimental context to deweight the
contribution of soft particles to event shapes, beyond the linear suppression in
$m_\perp$ necessary for infrared safety.

We have included a summary of some common universality classes in
\Tab{tab:classes}. For $n=0$ and $n=1$, the universality classes correspond
with two that appear for classic event shapes, so we will also refer to $g(r)=1$ as the
Jet Mass class and $g(r) = r$ as the E-scheme class.

In \Sec{sec:MC}, we will show that Pythia~8 and Herwig++ do indeed exhibit
universality when holding $n$ fixed but varying $a$, but as expected yield
different power correction parameters $\Omega_1^n$ when $n$ is varied.
\begin{table*}[t!]
\begin{tabular}{l|c|l}
Class & ~~~$g(r)$~~~  & ~~Event shape\\
\hline 
Jet Mass class ($\Omega_1^0$ or $\Omega_1^\rho$) & $1$
  & ~~$\rho_\pm$, $\tau_{2}$, $\tau^J$, $\tau_{(a)}^J$, $C^J$ 
  \\
E-scheme class ($\Omega_1^1$ or $\Omega_1^E$)~~~ &  $r$
  & ~~$\tau_{(a)}$, $\tau^E=\tau_2^E$, $C^E$, $\rho_\pm^E$, $\tau^R$, $\tau_{(a)}^R$, $C^R$, $\rho_\pm^R$
  \\
$r^n$ class ($\Omega_1^n$)& $r^n$ 
& ~~generalized angularities $\ang{n,a}$ in \eq{tna} \\
\hline
Thrust class ($\Omega_1^{g_\tau}$) & $g_\tau(r)$ 
& ~~$\tau$, $\rho_\pm^P$, $\tau_2^P$ \\
$C$-parameter class ($\Omega_1^{g_C}$)~~& $g_C(r)$
& ~~$C$ \\
$r^2$ class ($\Omega_1^2$)  & $r^2$ & ~~$\ang{2,a}$, $\tau_{(a\to-\infty)}^P$
\end{tabular}
\caption{Event shape classes with a universal first power correction parameter $\Omega_1^{g_e}$.
For a given event shape, the full power correction is $\Omega_1^e = c_e \Omega_1^{g_e}$.
\label{tab:classes}}
\end{table*}

\subsection{Classic Event Shapes and Mass Schemes}
\label{sec:classic_event_shapes}
In this subsection, we discuss $c_e$ and $g_e(r)$ for the more traditional event
shapes enumerated in \tab{classice}, and show how $g_e(r)$ changes when using
various measurement schemes for hadron masses.  Results for the corresponding
$f_e(r,y)$ were summarized above in \Tab{tab:f-functions}.  As already
mentioned, the results for $c_e$ are independent of the measurement scheme for
hadron masses, and can be computed directly with \eq{defncg}. For the various
classic event shapes they are summarized in \Tab{tab:cfunctions}.

\begin{figure}[t!]
\includegraphics[width=0.485\textwidth]{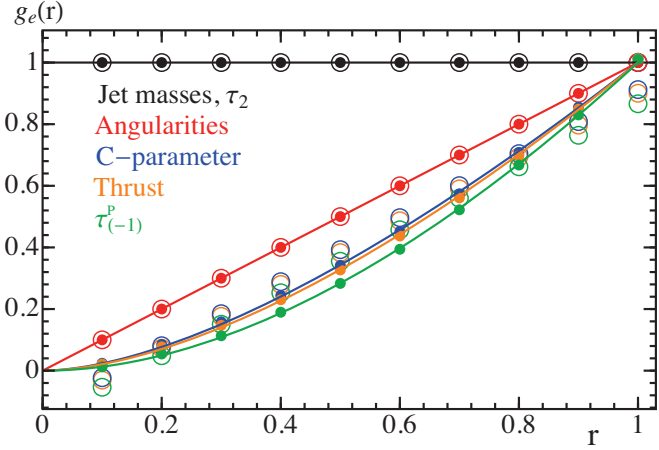}
\vspace{-0.5cm}
\caption{From top to bottom the $g_e(r)$ functions for Jet Masses and
  2-jettiness, Angularities, C-parameter, Thrust, and 
the P-scheme angularity with $a = -1$. Lines correspond to exact numerical 
values, open circles to two terms from the basis of \Sec{sec:basis}, and filled 
circles to three terms from the basis.
\label{fig:g-function}}
\end{figure}

For the event shapes in \tab{classice} with their original definitions,
integrating their $f_e(r,y)$ over $y$ we find
\begin{align} \label{eq:classicg}
  g_\tau(r) & = 1-E(1-r^2)+r^2\,K(1-r^2)  \,,\\
  g_{\tau_{2}}(r) = g_{\rho_\pm}(r) &= 1
  \,,\nn\\
  g_{\tau_{(a)}}(r) &= r
  \,,\nn\\
  g_C(r) &=\frac{2\,r^2}{\pi}K(1-r^2)
  \,,\nn
\end{align}
where $E$ and $K$ are the complete elliptic integrals
\begin{align}
 E(x)&= \!\int_0^{\pi/2} \!\!\!\!\!\! \df\theta\, (1\!-\!x\sin^2\theta)^{1/2},
  \\
 K(x)&= \!\int_0^{\pi/2} \!\!\!\!\!\! \df\theta\, (1\!-\!x\sin^2\theta)^{-1/2}
  \,. \nn
\end{align}
A plot of the $g(r)$'s in \eq{classicg} is displayed in
\Fig{fig:g-function}.

Direct analogs of $g_\tau(r)$, $g_C(r)$, $g_{\rho_\pm}(r)$ were computed in
\Ref{Salam:2001bd}. The translation from their notation (superscripts
${\rm SW}$) to ours is $c_e^{\rm SW}=c_e$ and $\delta c^{\rm SW}_e(m^2/p_\perp^2) = c_e [\,
g_e(r)/r -1]$. Fig.~1 of \Ref{Salam:2001bd} plots $(1+\delta
c_e^{\rm SW}/c_e^{\rm SW})$ versus $p_\perp/m$, which is the direct analog of our
\Fig{fig:g-function}. In our figure, the function is bounded because of our
use of the $r = p_\perp/m_\perp$ variable rather than $p_\perp/m$. These bounded $g(r)$ functions
are more convenient for our basis discussion in \Sec{sec:basis} below.

By looking for event shapes with common $g(r)$'s in \eq{classicg} or
\fig{g-function} we see that $\rho_\pm$ and $\tau_2$ are in the ``Jet
Mass'' universality class, and that the angularities $\tau_{(a)}$ for any $a$
belong to the ``E-scheme'' universality class.  Defining $\Omega_1^\rho$ and $\Omega_1^{E}$ as the universal
power corrections for the Jet Mass and E-scheme classes, respectively, and accounting for the $c_e$
factors, we have
\begin{align}
 \Omega_1^{\tau_2} &= 2\,\Omega_1^{\rho}   \,,
 &\Omega_1^{\tau_{(a)}} & = \frac{2}{1-a}\: \Omega_1^{E} \,.
\end{align}
Thrust and the C-parameter have their own $g(r)$ functions and hence among these event shapes
are alone in their universality classes, with power corrections
$\Omega_1^{\tau}$ and $\Omega_1^{C}$.\footnote{Accounting for the $c_e$ dependence, the universal power
corrections for the thrust and C-parameter classes are $\Omega_1^{g_\tau} = \Omega_1^{\tau}/2$ and
$\Omega_1^{g_C} = \Omega_1^{C}/3\pi$.}

We now consider the various hadron mass schemes.  For all of the event shapes in
\tab{classice} in the E-scheme, one has a common $g_e(r) = r$, so they all
belong to the ``E-scheme'' universality class.  It was for this reason that
\Ref{Salam:2001bd} considered the E-scheme to be privileged hadron mass scheme.
In this sense, universality in the E-scheme is closest to the universality for
massless particles, since any event shape that can written as a function of
four-vectors has an E-scheme definition and they all have the same universal
power correction. For event shapes defined directly in terms of $m_\perp$, $r$,
and $y$ (e.g.~the generalized angularities in \eq{tna}), the E-scheme can be
defined by first expressing the event shape in terms of four-vectors (with no
explicit $m$ dependence), and then applying the E-scheme replacement in
\Eq{eq:Escheme}.  Then any E-scheme observable has $g(r)=r$ and hence is in the
E-scheme class. An example of an exception are event shapes that vanish in the
massless limit, $g_e(r=1)=0$, which do not have a meaningful E-scheme definition.

From \Tab{tab:g-functions}, we see that the event shapes measured in the
R-scheme also fall into the E-scheme class \mbox{($g(r) = r$)}, while event shapes
measured in the J-scheme fall into the Jet Mass class ($g(r)=1$).  The fact that
E-scheme and R-scheme jet shapes are in a common universality class is a
non-trivial consequence of the relation $\partial \eta/\partial y = 1/v$ from
\Eq{eq:vrelations}.

For the P-scheme, universality classes are more complicated. Carrying out the integral in
\eq{defncg} to find the $g(r)$'s one finds that thrust and the C-parameter are
unchanged. For the jet masses and 2-jettiness \mbox{$ g_{\rho^P_\pm}(r) =
g_{\tau_2^P}(r) = g_{\tau}(r)$}, so both $\rho_\pm^P$ and $\tau_2^P$ belong to the
thrust universality class:
\begin{align}
   \Omega_1^{\rho_\pm^P}
    =\frac{1}{2} \Omega_1^{\tau_2^P} 
    =\frac{1}{2} \Omega_1^{\tau} \equiv \Omega_1^{g_\tau}
   \,. 
\end{align}
For P-scheme angularities, $g_{\tau^P_{(a)}}(r)$ does not appear to have a simple
analytic formula for arbitrary $a$ (although it is easy to compute numerically).
By making a change of variables $y = -\ln\tan(\chi/2)$, a convenient way of 
writing it is
\begin{align}
g_{\tau_{(a)}^P}(r) & = (1\!-\!a)\! \int_0^{\pi/2}\!\!\!\!\!\! \df \chi \
  \frac{\Big(\!\sqrt{1\!-\!(1\!-\!r^2)\sin^2\!\chi}+\cos \chi\!\Big)^{a-1}}
  {r^{a-2}\:\sin^{a}\! \chi}
  \,.
\end{align}
For integer values, it is simple to find an analytic form, and for $a=-1,\,-2$ we
find\footnote{In general for odd $a$, $g(r)$ involves only $\ln(1+r)$ and for even $a$
only the elliptic functions $K({1-r^2})$ and $E({1-r^2})$ appear.}
\begin{align}
g_{\tau^P_{(-1)}}(r) & = 2 - r -2\,\Big(\frac{1}{r}-r\Big)\ln(1+r)\,,\\
g_{\tau^P_{(-2)}}(r) &= 9-\frac{8}{r^2}-\Big(7-\frac{8}{r^2}\Big)\,E(1-r^2)\nn\\
& ~~~~~ -(4-3\,r^2)\,K(1-r^2)\,.\nn
\end{align}
For any $a$, $g_{\tau^P_{(a)}}(0)=0$ and $g_{\tau^P_{(a)}}(1)=1$. For large
negative values of $a$ one finds
\be
g_{\tau_{(a \to - \infty)}^P}(r) = r^2.
\ee
Hence angularities in the P-scheme for large negative $a$ belong to the same class as the generalized
angularities in \Eq{eq:tna} for $n=2$, with $\Omega_1^{\tau_{(a \to -\infty)}^P} = \Omega_1^{\ang{2,a}}$. 

In practice, $g_{\tau_{(a)}^P}(r)$ for arbitrary $a$ quickly converges
towards $g_{\tau^P_{(-\infty)}}(r)$, and hence there is a quasi-universality for
angularities in the P-scheme.  Also $g_{C}(r)$ and $g_{\tau}(r)$ are not so different from
$g_{\tau_{(-\infty)}^P}(r)$, implying an approximate universality between
event shapes in different classes for all event shapes in the P-scheme. This was
already noted for thrust and the C-parameter in \Ref{Salam:2001bd}.  We will next
develop a complete basis for describing $g_e(r)$ functions that will allow us to
make this observation more quantitative.

\subsection{Orthogonal Basis for $\boldsymbol{\Omega_1(r)}$}
\label{sec:basis}
\begin{table}[t!]
\begin{tabular}{l|cccc}
 & $b_0$ & $b_1$ & $b_2$ & $b_3$\\
\hline 
Jet masses, $\tau_2$ & $1$ & $0$ & $0$ & $0$\\
Angularities & $\dfrac{1}{2}$ & $\dfrac{1}{2\sqrt{3}}$ & $0$ & $0$\\
Thrust & ~~$0.383$~~ & ~~$0.299$~~ & ~~$0.050$~~ & ~~$-0.006$~~\\
$\tau_{(-1)}^{P}$ & $0.355$ & $0.295$ & $0.064$ & $-0.004$\\
C-parameter~~~~ & $0.393$ & $0.300$ & $0.046$ & $-0.007$\\
$\tau_{(a \to - \infty)}^{P}$ & $\dfrac{1}{3}$ & $\dfrac{1}{2\sqrt{3}}$ & $\dfrac{1}{6\sqrt{5}}$ & $0$\\
\end{tabular}\caption{Numerical value of the coefficients of the complete basis for the various $g_e(r)$ functions.\label{tab:cn}}
\end{table}
Many $g_e(r)$ curves are still parametrically close, even if the corresponding event shapes are formally in different
universality classes.  Examples are thrust, C-parameter, and $\tau_{(-1)}^P$ angularity shown in \Fig{fig:g-function}.
To get a quantitative handle on this observation, we can use a complete set of orthonormal functions
for $r\in [0,1]$:
\begin{align}
& h_n(r) \equiv \sqrt{2n+1}\,P_n(2r -1)\,, \label{eq:basis} \\
& \int_0^1 \df r\,h_n(r)\,h_m(r) = \delta_{nm}\,,\nn
\end{align}
where $P_n$ are the Legendre polynomials.  Now we can decompose any of the $g_e(r)$
functions in this basis in the usual way:
\begin{align}
g_e(r)& =\sum_{n=0}^\infty b^e_n \,h_n(r)\,,\label{eq:f-decomposition}   \\
b^e_n & = \int_0^1 \df r\,g_e(r)\,h_n(r)\,.\nn
\end{align}
Since the $g(r)$ functions for classic event shapes are fairly close to
low-order polynomials, the first few terms in the basis will provide an accurate approximation.

The values of the $b_n^e$ coefficients are shown in \Tab{tab:cn} for the classic
event shapes. The approximation for the $g_e(r)$ functions are plotted in
\Fig{fig:g-function}, where the exact values are shown by lines and the
approximation with two terms from the basis ($b_0$, $b_1$) are shown by hollow
circles, and with three terms ($b_0$, $b_1$, $b_2$) by filled circles.  For the
jet masses, 2-jettiness, and angularities, the approximate result is exact with
two terms in the basis. For the remaining events shapes, the approximation with
two terms is likely sufficiently accurate at the level one expects of current
experimental and perturbative precision.  With three terms, the approximation is
excellent in all cases, so the third term can be regarded as a high-precision
correction.  This is also apparent from the values in \Tab{tab:cn}, where $b_0$
and $b_1$ are much larger than $b_2$ (and computing $b_{n>2}$ one finds they are
negligible).

Using \Eq{eq:basis} one can write any $\Omega_1^e$ in terms of a denumerable set
of power correction parameters:
\be
\label{eq:omega-decomposition}
\Omega_1^e = \sum_{n=0}^\infty b_n^e\,\Omega_1^{(n)}\,, \qquad  \Omega_1^{(n)} = \int_0^1 \df r\, h_n(r)\,\Oop\,.
\ee
Only the first few terms will be numerically relevant for most event shape
observables. Using \Eqs{eq:f-decomposition}{eq:omega-decomposition} and the
results in \Tab{tab:cn}, we can make the following exact identifications
\begin{align}
\Omega_1^{(0)} = \Omega_1^\rho\,,\qquad
\Omega_1^{(1)} = 2\sqrt{3}\,\Omega_1^E-\sqrt{3}\,\Omega_1^\rho\,.
\end{align}
Thus the power correction parameters for the Jet Mass class and E-Scheme class
already give an excellent approximation for the various classic event shapes. To
refine the prediction even further one can use the next term in the basis,
$\Omega_1^{(2)}$.\footnote{In principle, we could extract $\Omega_1^{(2)}$
  exactly from the $\tau_{(a \to - \infty)}^{P}$ power correction, since from
  \Tab{tab:cn}, we see that $g_{\tau_{(a \to - \infty)}^{P}}(r)$ is saturated by
  the first three terms in the Legendre expansion.}  Writing the leading power
correction for the other event shapes in terms of $\Omega_1^\rho$, $\Omega_1^E$,
and $\Omega_1^{(2)}$ we have:
\begin{align}
\label{eq:PCrelations}
\Omega_1^\tau          
   & = 1.034\,\Omega_1^E - 0.135\,\Omega_1^\rho + 0.050\,\Omega_1^{(2)}
   \,,\\
\Omega_1^C             
   & = 1.039\,\Omega_1^E - 0.127\,\Omega_1^\rho + 0.046\,\Omega_1^{(2)}
   \,,\nn\\
\Omega_1^{\tau_{(-1)}^P} & = 1.022\,\Omega_1^E - 0.156\,\Omega_1^\rho 
  + 0.064\,\Omega_1^{(2)}
  \,.\nn
\end{align}
In general both the $\Omega_1^E$ and $\Omega_1^\rho$ terms are numerically
important since experiment favors values $\Omega_1^E\sim \Omega_1^\rho/2$, and
$\Omega_1^{(2)}$ can be typically neglected. However, since the numerical
coefficients in \eq{PCrelations} are so close, one is justified in using the
approximation $\Omega_1^\tau \simeq \Omega_1^C \simeq \Omega_1^{\tau_{-1}^P}$
up to corrections of $\sim 15\%$.

We note that the above basis analysis is valid only at a fixed value of $Q$
since $\Omega_1(r,\mu)$ has an anomalous dimension which we will compute in
\Sec{sec:running}, and the appropriate $\mu$ scales with $Q$.  As we will see in
\Sec{sec:newbasis}, we will need to refine the above analysis in order to relate
power corrections at different values of $Q$.

\subsection{Renormalon-free Definition of $\boldsymbol{\Omega_1(r)}$}
\label{sec:renormalon}
It is well known that the first moment of a perturbative event shape
distribution has a $\Lambda_{\rm QCD}$ renormalon ambiguity in the
$\overline{\rm MS}$ scheme~\cite{Gardi:2000yh}, which corresponds to a
renormalon in the $\overline{\rm MS}$ power correction parameter $\Omega_1^e$
that we have been considering so far. To our knowledge, the only
renormalon-based analysis of event shapes that observes sensitivity to hadron
masses is \Ref{Beneke:1998ui} (which in turn, called into question the massless
universality results).  In this section we argue that the universality relations
given in \Sec{sec:gclasses} will remain unchanged as long as one defines
appropriate renormalon-free schemes for the $1/Q$ power corrections.

In general the $\Lambda_{\rm QCD}$ renormalon is removed by converting the
power correction parameter to a new scheme
\begin{align} \label{eq:O1R}
  \Omega_1^e(R,\mu) &\equiv \Omega_1^e(\mu) - \delta_e(R,\mu)\,,
\end{align}
where $\delta_e(R,\mu)$ is a series in $\alpha_s(\mu)$. There is a corresponding
change to the perturbative part of the cross section that depends on this same
series. Writing $\hat\sigma_e(x)$ for the Fourier transform of
$\df\hat\sigma/\df e$ this change is
\begin{align} \label{eq:sigmay}
  \hat\sigma_e(x) \to \tilde\sigma_e(x)=\hat\sigma_e(x)\: e^{-ix\,
    \delta_e(R,\mu)/Q} \,.
\end{align}
Recall that universality classes relate $\Omega_1^e(\mu)$ for
different event shapes $e$, and that the relations are nonperturbative. Hence it
is clear that one has the \emph{same} renormalon ambiguity for event shapes that
are members of the same class (differing only by the proportionality
constants $c_e$). Thus it is sufficient to adopt a common scheme change for
members of the same class via any scheme satisfying
\begin{align} \label{eq:deltaseries}
  \delta_e(R,\mu) &= c_e\: R\, e^{\gamma_E}  \sum_{n=1}^\infty 
       \Big(\frac{\alpha_s(\mu)}{4\pi}\Big)^n 
  \delta_n^{(g_{e'})}(\mu/R)
   \,.
\end{align}
Here the $\delta_n^{(g_{e'})}(\mu/R)$ coefficients involve factors of
$\ln(\mu/R)$ and are chosen such that the new $\Omega_1^e(R,\mu)$ is free of the
leading $\Lambda_{\rm QCD}$ renormalon ambiguity. The numerical values for
$\delta_n^{(g_{e'})}$ will depend on the representative $e'$ that we choose, but
any representative has the same renormalon and hence is a valid subtraction
series for all members of the $g_e$ class.

One popular scheme for fixing the coefficients $\delta_n^{(g_{e'})}(\mu/R)$ is
based on the dispersive approach, where $R=\mu_I$~\cite{Dokshitzer:1995zt,
  Dokshitzer:1995qm,Dokshitzer:1998pt}. In this framework to obtain an
appropriate subtraction at ${\cal O}(\alpha_s^2)$ one must also include the
Milan factor~\cite{Dokshitzer:1997iz,Dokshitzer:1998pt}. As is typical
of scheme changes based on QCD perturbation theory, this scheme removes the
renormalon for the massless limit $r=1$.

At higher orders in $\alpha_s$ a more convenient scheme uses the leading power
perturbative soft function itself~\cite{Hoang:2007vb}, $S_{e}^{\rm pert}$, which
does not require additional computations beyond those needed for a resummed
analysis of the event shape.  Again this scheme change removes the renormalon
for the massless limit $r=1$. For each $g_e$ universality class we pick a
representative $e'$ that is a member of the class.  Then for all event shapes
$e\in g_e$ we define\footnote{ To facilitate a multiplicative renormalization
  structure and to account for non-Abelian exponentiation, one uses the
  logarithmic derivative of the position space soft function
  $S^\pert_e(x,\mu)$~\cite{Hoang:2008fs}, a definition adapted from the top
  jet-mass definition of \Ref{Jain:2008gb}.}
\begin{align} \label{eq:deltadefn}
  \delta_e(R,\mu) = \frac{c_e}{c_{e'}}\, R e^{\gamma_E} \frac{\df}{\df\ln(ix)}
  \ln S^\pert_{e'}(x,\mu) \bigg|_{x=(iRe^{\gamma_E})^{-1}}\,. 
\end{align}
For $e=\tau$ \Eq{eq:deltadefn} yields the subtractions for thrust defined in
\Ref{Hoang:2008fs}. Given that renormalons probe the infrared structure of
amplitudes, one might naively expect that different subtractions would be necessary
for the different universality classes, since they treat hadron masses
differently.  However, the scheme change in \eq{deltadefn} is based solely on the QCD
perturbative calculation of $S_{e'}^{\rm pert}$ with $r=1$, so only the $c_e$
coefficients will differ in subtractions for different event shapes. Thus in this
setup, the same subtraction $\delta_{e'}(R,\mu)$ can be chosen for
all universality classes.

In \App{app:renormalon} we carry out a standard renormalon bubble sum
calculation for an arbitrary dijet event shape $e$ satisfying \eq{f-definition}
and demonstrate that \eq{deltadefn} yields a perturbative cross section $\tilde
\sigma_e(x)$ and power correction parameter $\Omega_1(R,\mu)$ that are free from
the $\Lambda_{\rm QCD}$ renormalon probed by this method. It might be
interesting to consider extensions of \eq{deltadefn} that satisfy
\eq{deltaseries} but have additional dependence on hadron mass effects.

\section{Anomalous Dimension of $\boldsymbol{\Omega_1(r)}$}
\label{sec:running}
\subsection{Running at One-Loop}
\label{subsec:onelooprun}
In this section, we compute the one-loop anomalous dimension of the QCD matrix
element $\Omega_1(r,\mu)$ in the $\overline{\rm MS}$ scheme, with details given
in \App{app:anomdim}.  We regulate the UV with dimensional regularization (DR)
using $d=4-2\epsilon$.  

Since $\Omega_1(r)$ has
mass-dimension one and is proportional to the infrared (IR) scale $\Lambda_{\rm QCD}$, one must
be careful to establish IR regulators for the perturbative calculation in such a way
that there is nonzero overlap with the operator matrix element. This requires at
least one dimensionful IR regulator, as well as a mechanism to probe different values
of $r$. With this, we can then compute the anomalous dimension just as we would
for any external operator in QCD.  The anomalous dimension will be independent
of the precise IR procedure used to identify the matrix element.  
 
A  convenient choice for the IR regulator is obtained by coupling a massive adjoint
background source $J^{\mu A}$ to the Wilson lines in $\Omega_1(r)$ by the
replacement
\begin{align}
  A^{\mu A}(x) \to A^{\mu A}(x) + J^{\mu A}(x)\,,
\end{align}
in \eq{wilson-lie} and in the QCD Lagrangian. There is no Lagrangian mass term
for the source $J^{\mu A}$, but for it to serve as an IR regulator, we will consider it
to carry a massive particle momentum $q^\mu$ where $q^2=m^2$.
Recall from \eq{trans-mom-flow-op} that the ${\cal E}_T(r,0)$ operator in $\Omega_1(r)$ sums
over contributions from individual particles in the final state. If ${\cal E}_T(r,0)$ acts on any
particle other than $J^{\mu A}$, then the corresponding
phase space integral is scaleless and dimensionful, and hence vanishes in DR.
Thus the first nonzero contribution occurs where ${\cal E}_T(r,0)$ acts on a
$J^{\mu A}$ (and we then set all other $J^{\mu A}$'s to zero). This IR regulator
is convenient for bookkeeping and does not overly complicate the evaluation of
loop integrals. Effectively it amounts to considering one of the final state gluons,
namely the one acted upon by ${\cal E}_T(r,0)$, as having mass $m$ and thus $r\ne 1$.
For convenience when drawing Feynman diagrams, we use the same notation for
massless gluons and the source, and simply note that we must sum over the cases
where each final state gluon is the source.

\begin{figure}[t!]
\includegraphics[width=0.35\textwidth]{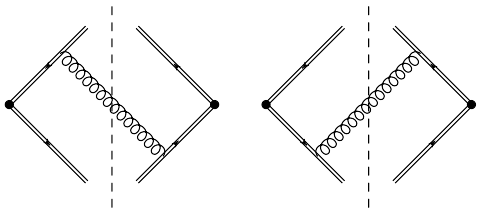}
\caption{Tree level graphs for $\Omega_1(r,\mu)$. 
\label{fig:tree}}
\end{figure}

At tree level, as shown in \Fig{fig:tree}, we have only the source line. This yields the
nonzero matrix element
\begin{equation}\label{eq:tree-level-c}
M_1^{\rm tree}(r) =
\frac{2\alpha_sC_F}{\pi}\,\frac{m\,r}{(1-r^2)^{\frac{3}{2}}}
 \,.
\end{equation}
Hence our setup provides nonzero overlap with the operator in $\Omega_1(r)$. 

We now turn our attention to the ${\cal O}(\alpha_s^2)$ corrections.  Here we
must fix a gauge for the $A^{\mu A}$ massless gluons.  We have carried out all
our calculations both with traditional Feynman gauge, as well as with a
background field Feynman gauge where the source $J^{\mu A}$ takes the place of
the external background field.  Both yield the same results.  In addition to $J^{\mu A}$, we
must introduce extra IR regulators specific to individual diagrams. We find that
shifting eikonal propagators involving the loop or phase-space momentum
$k^\mu$ by $n\cdot k \to n\cdot k + \Delta_n$ and $\bar n\cdot k\to \bar n\cdot
k+\Delta_{\bar n}$ suffices to regulate other IR divergences.

\begin{figure}[t!]
\includegraphics[width=0.35\textwidth]{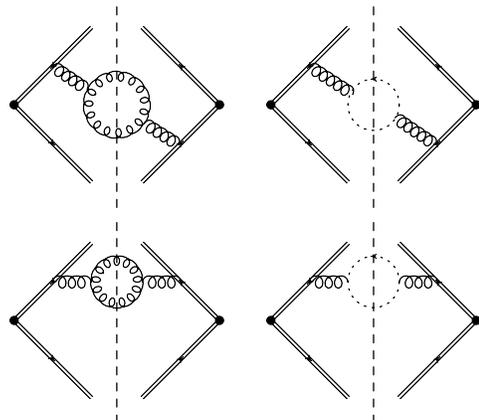}
\vspace{-0.5cm}
\caption{Diagrams involving gluon or ghost bubbles with two cut
  particles. The 4 additional diagrams obtained by a flip about the horizontal
  or vertical axis are not shown.  Diagrams with one cut particle for
  wavefunction renormalization and coupling renormalization are also not
  displayed.
\label{fig:self-energy}}
\end{figure}

For the diagrams in \Fig{fig:self-energy} which involve two particles in the final state through
either a ghost bubble, or a gluon bubble, the sum of graphs does not involve a UV divergence
(moreover these graphs are IR finite). Hence they can be ignored for the purposes of calculating
the anomalous dimension.
\begin{figure}[t!]
\includegraphics[width=0.35\textwidth]{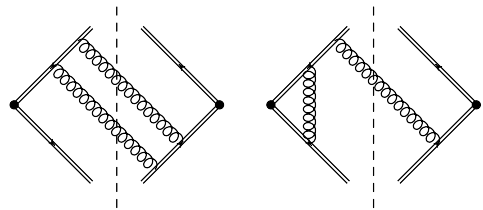}
\vspace{-0.3cm}
\caption{Purely Abelian ${\cal O}(\alpha_s^2)$ diagrams. 
  Either gluon line crossing the cut can be the source.  The 4 additional
  diagrams obtained by a flip about the horizontal or vertical axis are not
  shown.
\label{fig:abelian}}
\end{figure}
The remaining diagrams are either Abelian with color factor $C_F^2$ or
non-Abelian with color factor $C_F C_A$. Abelian contributions are shown in
\Figs{fig:abelian}{fig:non-abelian}, and in the sum over Abelian diagrams, the real
radiation and virtual contributions exactly cancel (for all $\epsilon$). It is easy
to prove that this cancellation happens to all orders in perturbation
theory. This proof uses the fact that $Y[A+J]=Y[A]Y[J]$ for an Abelian
theory with no light quarks, and that $Y[A] Y^\dagger[A]=1$.

\begin{figure}[t!]
\includegraphics[width=0.49\textwidth]{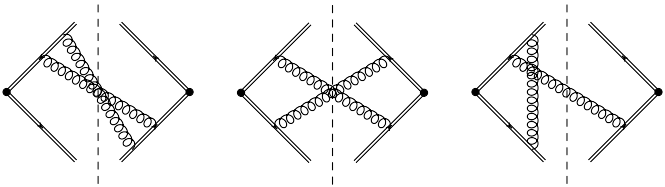}
\vspace{-0.5cm}
\caption{Independent emission diagrams with Abelian and non-Abelian
contributions.  The 4 additional diagrams obtained by a horizontal flip 
or complex conjugation are not shown. 
  \label{fig:non-abelian}}
\end{figure}

\begin{figure}[t!]
\includegraphics[width=0.35\textwidth]{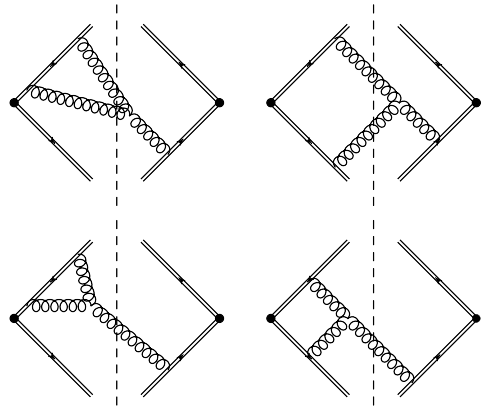}
\vspace{-0.5cm}
\caption{Triple gluon Y-diagrams for the ${\cal O}(\alpha_s^2)$ correction 
to $\Omega_1(r)$. 
The 12 additional diagrams obtained by a horizontal flip 
or complex conjugation are not shown. Diagrams with all 3 gluons coupled to Wilson
lines of the same direction vanish.
\label{fig:triple-gluon}}
\end{figure}

This leaves just the non-Abelian diagrams coming from \Fig{fig:non-abelian}, and
triple gluon vertex diagrams in \Fig{fig:triple-gluon}. There is also a
contribution from gauge coupling renormalization which is not just canceled by
the vacuum polarization graphs in many gauges.  For the $1/\epsilon$ poles, we
can take $\Delta_{n,\bar n}\to 0$ in the sum of graphs in \Fig{fig:non-abelian}
and separately in \Fig{fig:triple-gluon}, so the extra IR regulators cancel out
of the UV terms as expected.  The sum of diagrams in \Fig{fig:non-abelian} and
the sum in \Fig{fig:triple-gluon} each have $1/\epsilon^2$ poles, but these
cancel in the complete sum. This leaves only a nonzero $1/\epsilon$ pole which
will yield the anomalous dimension. 

The final result for the UV divergence in the matrix element of the bare
$\Omega_1(r)$ operator is
\begin{equation} \label{eq:M1bare}
 M_1^{1\text{-loop}}(r) = \Big(-\frac{\alpha_s C_A}{2\pi\epsilon}\ln(1-r^2)\Big) \
 M_1^{\rm tree}(r)\,.
\end{equation}
We define the renormalized $\overline{\rm MS}$ operator by
\begin{align}
 \Omega_1^\text{bare}(r) &= Z(r,\mu,\epsilon)\ \Omega_1(r,\mu) \,,
\end{align}
so \eq{M1bare} determines $Z(r,\mu,\epsilon)$ at ${\cal O}(\alpha_s)$.  Using
$\mu \, \df\alpha_s/\df\mu = -\,2\,\epsilon\,\alpha_s+\ldots\,$, the
one-loop anomalous dimension of $\Omega_1(r)$ is
\begin{equation}\label{eq:O1-anom-dim}
\mu\,\frac{\df }{\df \mu}\Omega_1(r,\mu) = \Big[-\frac{\alpha_s C_A}{\pi}\ln(1-r^2)
\Big] \ \Omega_1(r,\mu) \,.
\end{equation}
Note that this anomalous dimension is positive since $\ln(1-r^2)<0$.

Intriguingly, the anomalous dimension is $r$-dependent, showing the important role of hadron masses.
However, there is no mixing for operators at different values of $r$, so this Renormalization Group Evolution
equation can be solved exactly to yield
\begin{align} \label{eq:fullRGE}
\Omega_1(r, \mu) 
  &= \Omega_1(r, \mu_0) \:\bigg[
  \frac{\alpha_s(\mu)}{\alpha_s(\mu_0)}\bigg]^{
  \frac{2\,C_A}{\beta_0}\ln(1-r^2)} \\
  &= \Omega_1(r, \mu_0) \big[ 1-r^2 \big]^{
  \frac{2\,C_A}{\beta_0}\ln\frac{\alpha_s(\mu)}{\alpha_s(\mu_0)} }
  \nn \,.
\end{align}
Here one can consider $\mu_0\sim 2~\GeV$ as the low energy hadronic scale where
we specify the nonperturbative matrix element, and $\mu$ as a high energy scale
that is appropriate for the observable being considered.  In \Sec{sec:ope} we
will show that $\mu\simeq Qe$ for the region of event shape distributions with
$e\ll e_{\max}$ and $Qe\gg \Lambda_{\rm QCD}$, and $\mu\simeq Q e_{\max}$ for
first moments of event shapes.

In order to use the resummed expression for $\Omega_1(r,\mu)$ to predict the
evolution from $\Omega_1^e(\mu_0)$ to $\Omega_1^e(\mu)$, one would need to know the full
$r$-dependence of $\Omega_1(r,\mu_0)$ to perform the integral over $g_e(r)$.
We will see how to approximately circumvent this problem in \Sec{sec:newbasis}.

One can also consider expanding \eq{fullRGE} perturbatively in $\alpha_s(\mu_0)$ which yields
\begin{align}
\Omega_1(r, \mu) &= \Omega_1(r,\mu_0) \\
& ~~~ \times \Big[1-\frac{\alpha_s(\mu_0)C_A}{\pi}\ln\Big(\frac{\mu}{\mu_0}\Big)\ln(1-r^2)+\ldots \Big]\,. \nn
\end{align}
If one truncates at ${\cal O}(\alpha_s)$, then one only needs two nonperturbative parameters defined at $\mu_0$
to determine the the power corrections for an event shape at a higher scale $\mu$:
\begin{align} \label{eq:expandedrunning}
\Omega_1^e(\mu) & = \Omega^e_1(\mu_0)+\frac{\alpha_s(\mu_0) C_A}{\pi}
\ln\Big(\frac{\mu}{\mu_0}\Big)\, \Omega^{e,\,\ln}_1(\mu_0)\,.
\end{align}
Here $\Omega^e_1(\mu_0)$ is our standard power correction parameter at the
scale $\mu_0$ given by \eq{universality} with $\Omega_1(r,\mu_0)$, and the slope parameter $\Omega^{e,\,\ln}_1(\mu_0)$ is defined by
\begin{align} \label{eq:Oln}
\Omega^{e,\,\ln}_1(\mu_0) &\equiv - \!\int\! \df r \, \ln(1-r^2)\,
  c_e\,g_e(r)\,\Omega_1(r,\mu_0)\,.
\end{align}
The expanded form in \eq{expandedrunning} is a reasonable approximation if $\mu$
is not too different from $\mu_0$. It works for a larger range than one would
naively expect since there are numerical cancellations in the ${\cal
  O}(\alpha_s^2)$ term between $\ln^2(1-r)$ and $\ln(1-r)$ contributions.

\subsection{The Wilson Coefficient of $\boldsymbol{\Omega_1(r,\mu)}$}
\label{sec:ope}
Having established that the nonperturbative matrix element $\Omega_1(r,\mu_0)$ runs, 
we reconsider the operator expansion of the shape function
$F_e(\ell)$ in \eq{shape-function}, now incorporating $\alpha_s$ corrections
through a Wilson coefficient $C_1^e(\ell,r,\mu)$ for $\Omega_1(r,\mu)$.  The
formula in \eq{shape-function} becomes
\begin{align}\label{eq:shape-function2}
F_e(\ell) &= \delta(\ell) + \int\! \df r\: C_1^e(\ell,r,\mu)\:
  c_e\, g_e(r) \,\Omega_1(r,\mu) \nn\\
 & ~~~ + {\cal O}\Big(\frac{\Lambda_{\rm QCD}^2}{\ell^3}\Big) \,.
\end{align}
As usual the $\mu$ dependence of $C_1^e(\ell,r,\mu)$ cancels that of
$\Omega_1(r,\mu)$. The dependence of $C_1^e(\ell,r,\mu)$ on $\ell$ and $\mu$
will determine the appropriate scale $\mu$ where there are no large logarithms in this
Wilson coefficient.  This in turn will determine the appropriate perturbative
scale $\mu$ for the endpoint of the evolution derived in \eq{fullRGE}.  Since the
$\ell$ dependence is treated differently by event shape distributions and by their first
moments, a different scale $\mu$ will be found for these two observables.

Taking \Eq{eq:O1-anom-dim} together with the
cancellation of the $\mu$ dependence implies
\begin{align}
  \mu \, \frac{\df}{\df\mu} \, C_1^e(\ell,r,\mu) 
   &= \frac{C_A\alpha_s(\mu)}{\pi} \ln(1\!-\! r^2)\, C_1^e(\ell,r,\mu) \,.
\end{align}
At order $\alpha_s$ using \Eq{eq:shape-function} this becomes
\begin{align}
  \mu \, \frac{\df}{\df\mu} \, C_1^e(\ell,r,\mu) 
   &= -\,\frac{C_A\alpha_s(\mu)}{\pi} \ln(1\!-\! r^2)\, \delta'(\ell).
\end{align}
Note that $C_1^e(\ell,r,\mu)$ must have mass dimension $-2$.  At ${\cal
  O}(\alpha_s)$ the simplest potential solution has the dependence
$\ln(\mu/\kappa)\delta'(\ell)$, but by dimensional analysis 
the only possibility for $\kappa$ is $\ell$
which leads to a singular result. The correct solution is
\begin{align} \label{eq:C1e}
C_1^e(\ell,r,\mu) &= -\,\delta^{\,\prime}(\ell) 
  + \frac{C_A\alpha_s(\mu)}{\pi} \ln(1\!-\! r^2) \frac{\df}{\df\ell} \left(
   \frac{1}{\mu} \Big[\frac{\mu}{\ell}\Big]_+ \right)
   \nn\\
 &~~~
  + \frac{\alpha_s(\mu)}{\pi} \, \delta^{\,\prime}(\ell)\, d_1^e(r) 
  +{\cal O}(\alpha_s^2) \,,
\end{align}
which can be deduced since the derivative of the plus function has the right
dimension and has the required logarithmic scale dependence
\begin{align}
 \mu\frac{\df}{\df\mu}\: \frac{\df}{\df\ell}
   \frac{1}{\mu} \Big[\frac{\mu}{\ell}\Big]_+ 
 &= -\,\delta^{\,\prime}(\ell) \,.
\end{align}
In this way, the plus function term in the Wilson coefficient exactly
compensates for the first order $\alpha_s(\mu) \ln(\mu)$ dependence in
$\Omega_1(r,\mu)$. Note that
\begin{align} \label{eq:plusderiv}
   \frac{\df}{\df\ell} \frac{1}{\mu} \Big[\frac{\mu}{\ell}\Big]_+ 
  &= - \frac{1}{\mu^2} \Big[\frac{\mu^2}{\ell^2}\Big]_{++} 
     + \frac{1}{\mu} \,\delta(\ell)\,,
\end{align}
where the $++$-distribution induces two subtractions about $\ell=0$ and is 
defined so that its zeroth and first moments integrate to zero for the limits
$\ell/\mu\in [0,1]$.

The function $d_1^e(r)$ in \eq{C1e} is also a perturbatively computable
contribution to the Wilson coefficient.  The matching calculation for this term
involves considering the difference between renormalized Feynman diagrams for the
full theory soft function matrix element
\begin{align}
 S_e(\ell) = \langle \, 0\,|\, \overline{Y}_{\bar n}^\dagger Y_n^\dagger
    \delta(\ell- Q\hat e) Y_n \overline{Y}_{\bar n}\, |\, 0\,\rangle\,,
\end{align}
and for the low-energy matrix elements describing $\Omega_1(r,\mu)$. A complete
one-loop calculation of $d_1^e(r)$ is beyond the scope of our work. In
\App{app:matching} we carry out this matching procedure for thrust
in order to directly derive the term that involves the derivative of the
plus-function shown in \eq{C1e}.  Many of the complications required to
derive $d_1^e(r)$ do not enter for this term.

Next consider the impact of $C_1^e(\ell,r,\mu)$ on the distribution and first
moment event shape observables discussed in \Sec{sec:observables}, in order to
determine the appropriate scale $\mu$ where large logs are minimized.  For the
differential distribution we find
\begin{align} \label{eq:distnpert}
  \frac{\df\sigma}{\df e} &= \frac{\df\hat\sigma}{\df e} 
  + \frac{1}{Q} \int\! \df\ell\! \int\!\df r\: \frac{\df\hat\sigma}{\df e}
    \Big(e-\frac{\ell}{Q}\Big) C_1^e(\ell,r,\mu) 
   \nn\\
  &\qquad \times c_e\, g_e(r)\, \Omega_1(r,\mu) 
   \nn\\[4pt]
 &= \frac{\df\hat\sigma}{\df e} 
  - \frac{1}{Q}\Big(\Omega_1^e(\mu) 
  +\frac{\alpha_s(\mu)}{\pi}\,\Omega_1^{e,\,d}(\mu) \Big) 
  \frac{\df^2\hat\sigma}{\df e^2}(e)
  \nn \\
  &\qquad + \frac{\Omega_1^{e,\,\ln}(\mu)}{Q} \frac{\alpha_s(\mu)C_A}{\pi}
   \Bigg\{ 
  \ln\Big(\frac{\mu}{eQ} \Big) \frac{\df^2\hat\sigma}{\df e^2}(e)
  \nn\\
  &\qquad - \int_0^{eQ}\!\! \frac{\df\ell}{\ell} \bigg[ 
   \frac{\df^2\hat\sigma}{\df e^2}\Big(e-\frac{\ell}{Q}\Big)-
  \frac{\df^2\hat\sigma}{\df e^2}(e) \bigg] 
  \Bigg\}
  \,,
\end{align}
where the nonperturbative parameter $\Omega_1^{e,\,\ln}(\mu)$ is given in
\eq{Oln}, and
\begin{align}
\Omega_1^{e,\,d_1}(\mu) = \int\! \df r\, d_1^e(r)\, c_e\, g_e(r) \, \Omega_1(r,\mu)\,.
\end{align}
The explicit $\ln(\mu/Qe)$ in \eq{distnpert} implies
that for the distribution, the appropriate scale to run the power correction to is $\mu=Qe$.

For the first moment we find
\begin{align} \label{eq:M1pert}
  \langle e\rangle &= \int_0^{e_{\rm max}}\!\!\!\!\df e\: e \int\!\df\ell\:
   \frac{1}{\hat\sigma}\frac{\df\hat\sigma}{\df e}\Big(e-\frac{\ell}{Q}\Big)
   \,F_e(\ell) 
   \\
  &= \int\! \df e \!\int\! \df\ell\:
  \theta\Big(e_m\!-\!e \!-\!\frac{\ell}{Q}\Big)\: \Big(e+\frac{\ell}{Q}\Big)
  \frac{1}{\hat\sigma}\frac{\df\hat\sigma}{\df e}(e)\: F_e(\ell) 
  \nn\\
 &= \vev{e}_\pert + \frac{\Omega_1^e(\mu)}{Q} 
 +\frac{\alpha_s(\mu)}{\pi}\,\frac{\Omega_1^{e,\,d}}{Q}
   + \frac{\Omega_1^{e,\,\ln}(\mu)}{Q} \frac{C_A\alpha_s(\mu)}{\pi}
  \nn\\
  &\times\!\! \int_0^{e_{\rm max}} \!\!\!\!\!\! \df e 
   \frac{1}{\hat\sigma}\frac{\df\hat\sigma}{\df e}(e)
   \bigg[ \ln\Big(\frac{\mu}{Q(e_{\rm max}\!-\!e)}\Big)
    - \frac{e^2}{e_{\rm max}(e_{\rm max}\!-\!e)}\bigg]
  \nn
\end{align}
where the notation $\vev{e}_\pert$ is defined in \Eq{eq:defpertmoment}.
Since the perturbative moments generate a rapidly convergent series, we can
expand the perturbative coefficient of $\Omega_1^{e,\,\ln}(\mu)$ in $(e/e_m)$ to
obtain
\begin{align} \label{eq:muchoice}
 \int_0^{e_{\rm max}} \!\!\! \df e &
   \frac{1}{\hat\sigma}\frac{\df\hat\sigma}{\df e}(e)
   \bigg[ \ln\Big(\frac{\mu}{Q(e_{\rm max}\!-\!e)}\Big)
   - \frac{e^2}{e_{\rm max}(e_{\rm max}-e)}\bigg]
  \nn\\
 &= \ln\Big(\frac{\mu}{Qe_{\rm max}}\Big) + \frac{\vev{e}_\pert}{e_{\rm max}} 
  -\frac{\vev{e^2}_\pert}{2 e_{\rm max}^2} + \ldots
\end{align}
Thus for the first moment the appropriate scale to run the power correction to
is $\mu=Q e_{\rm max}$.

\subsection{Orthogonal Basis  for $\boldsymbol{\Omega_1(r,\mu)}$ }
\label{sec:newbasis}
In \Sec{sec:basis}, we showed that the power corrections for observables in
different universality classes could be approximately related by expanding out
$g_e(r)$ and $\Omega_1(r,\mu)$ in a suitable basis.  Since the appropriate scale
$\mu$ for $\Omega_1$ is $Q$ dependent, this is true if all measurements are
performed at a single $Q$.  From \Secs{subsec:onelooprun}{sec:ope}, we know that
the leading power correction has nontrivial $Q$ dependence, and we would like to
incorporate this information in our description of $\Omega_1(r)$.

Immediately from \Eq{eq:fullRGE}, we see that $\Omega_1(r,\mu)$ will diverge as
$r\to 1$.  Even if $\Omega_1(r,\mu_0)$ is regular at some scale $\mu_0$, it will
quickly develop a singularity at $r = 1$ as $\mu$ evolves.  This singularity can
be physically interpreted as the propensity of Wilson lines to emit soft
massless particles.  Of course, this singularity is still square-integrable, and
thus the power correction is still well-defined.\footnote{For evolution over a
  large enough $Q$ range, the singularity may no longer be integrable, and may
  turn into a distribution, but we do not encounter this subtlety in our
  analysis.} However the singularity at $r = 1$ means that if the Legendre
polynomial basis in \Sec{sec:basis} is used at a scale $\mu_0$ to define a basis
for power corrections $\Omega_1^{(n)}(\mu_0)$, then it converges very slowly
when trying to use these same parameters to describe power corrections for scales
$\mu\gg \mu_0$ or $\mu \ll \mu_0$.

Looking at the $\alpha_s$ expansion of the running formula in
\Eq{eq:expandedrunning}, we see that to describe the power correction for a
range of $Q$ values, we must find a suitable basis to describe not only $g_e(r)$
and $\Omega_1(r,\mu_0)$ but also $g_e(r) \ln(1-r^2)$.  Since $\ln(1-r^2)$ is
unbounded at $r = 1$ (but still square integrable), we will use the
freedom to introduce additional square-integrable basis elements $(1-r)^{-k}$
for suitable values $0<k < 1/2$.  As long as we are content to work with a
finite number of basis elements, we can always make such a basis orthonormal via
the Gram-Schmidt procedure.  Adding these additional functional forms
will in general yield an over-complete basis, but this is not an issue in practice
since we will only ever consider a finite number of basis elements.

The situation becomes a bit more complicated if we consider the full running in
\Eq{eq:fullRGE}.  In general, any finite basis we choose to describe
$\Omega_1(r,\mu)$ at one value of $Q$ will not provide a good description at a
different value of $Q$ due to the running.  In particular, a basis that is
orthonormal at one value of $\mu_0$ will no longer be orthonormal at another
scale. Instead of trying to find a basis that works for any $Q$, we instead
choose a scale $\mu_0$ at which we model $\Omega_1(r,\mu_0)$, and then evolve
according to \Eq{eq:fullRGE}.  We then fit for the basis coefficients by using
information at different values of $Q$.  This procedure is philosophically the
same as the procedure used to determine parton distribution functions (PDFs),
where the PDFs are modeled at a low scale and then evolved to higher $Q$ values
via the Dokshizer-Gribov-Lipatov-Altarelli-Parisi equations~\cite{Dokshitzer:1977sg,Gribov:1972ri,Altarelli:1977zs}.

\subsection{Comparison to Monte Carlo}
\label{sec:MC}
We now show that power correction universality and running is exhibited by two
widely used Monte Carlo programs: Pythia~8.162 \cite{Sjostrand:2007gs} and
Herwig++ 2.6.0 \cite{Bahr:2008pv}. The hadronization model in Pythia~8 is based
on string fragmentation while Herwig++ is based on cluster fragmentation.\footnote{The
two programs also have different showering models, with Pythia~8
using a $p_\perp$-ordered shower and Herwig++ using an angular-ordered shower. 
Since the showers are evolved down to the non-perturbative scale, part of the
renormalization group evolution of the power correction may be captured by the
showering algorithm, and not just the hadronization model.}  The
default hadronization parameters in both programs have been tuned to reproduce
LEP $e^+ e^-$ event shapes at the $Z$ pole.  We will consider $e^+ e^- \to
\text{hadrons}$ at various $Q$ values, turning off initial-state electromagnetic
radiation to avoid the radiative return process.

Our study will use the generalized angularities $\ang{n,a}$ defined in
\Eq{eq:tna}.  Via the arguments in \Sec{sec:toy-event-shapes}, we know that the
power corrections for different values of $a$ are related via the $c_a =
2/(1-a)$ coefficients in \Eq{eq:cadef}, 
\be 
 \Omega_1^{\ang{n,a}}(\mu) = c_a\, \Omega_1^n(\mu),
\ee
where $\Omega_1^n(\mu)$ is the universal power correction for the $r^n$ class
given in \Eq{eq:universalrn}.

From \Eq{eq:momshift}, we know that the leading power correction shifts the
moments of event shapes, so we can extract the universal power corrections for
the generalized angularities via the first moment
\begin{equation}
 \Omega_1^n(\mu_Q)  =  \frac{1}{c_a} \left(Q \vev{\ang{n,a}}  - Q  \vev{\ang{n,a}}_\pert \right), 
\end{equation}
up to small higher-order corrections.  Since the maximum value of thrust
($a=0$) is $1/2$, we take $\mu_Q=Q/2$ to avoid having large logs in these
higher order corrections, which  were displayed above in
\Eq{eq:muchoice}.\footnote{The angularities with $a \not= 1$ have different
  values for $e_{\rm max}$, but this is an ${\cal O}(1)$ change to the $\mu$
  scale, and hence not relevant.}  The perturbative moment
$\vev{\ang{n,a}}_\pert$ is the same for event shapes with a common $r \to 1$
limit.  To form a combination that is sensitive to the power corrections in the
Monte Carlo programs without having to know about their perturbative
contributions, we consider the difference $\ang{0,a}-\ang{n,a}$ which compares
the same value of $a$ at two different values of $n$. For this combination we
have
\be 
{\Omega}_1^0(\mu_Q) - {\Omega}_1^n(\mu_Q) = \frac{Q}{c_a}\, \big( \vev{\ang{0,a}} -
\vev{\ang{n,a}}\big).  
\ee
Note that this difference is also independent of the additive scheme change that
removes the renormalon from $\Omega_1^e(\mu)$, given by \eq{O1R} with
\eq{deltadefn}. Therefore we emphasize that our analysis in this section only probes
the running in \eq{O1-anom-dim} and not the
$R$-evolution~\cite{Hoang:2008yj,Hoang:2009yr} associated with
$\Omega_1^e(R,\mu)$.
\begin{figure}[t!]
\includegraphics[width=0.8\columnwidth]{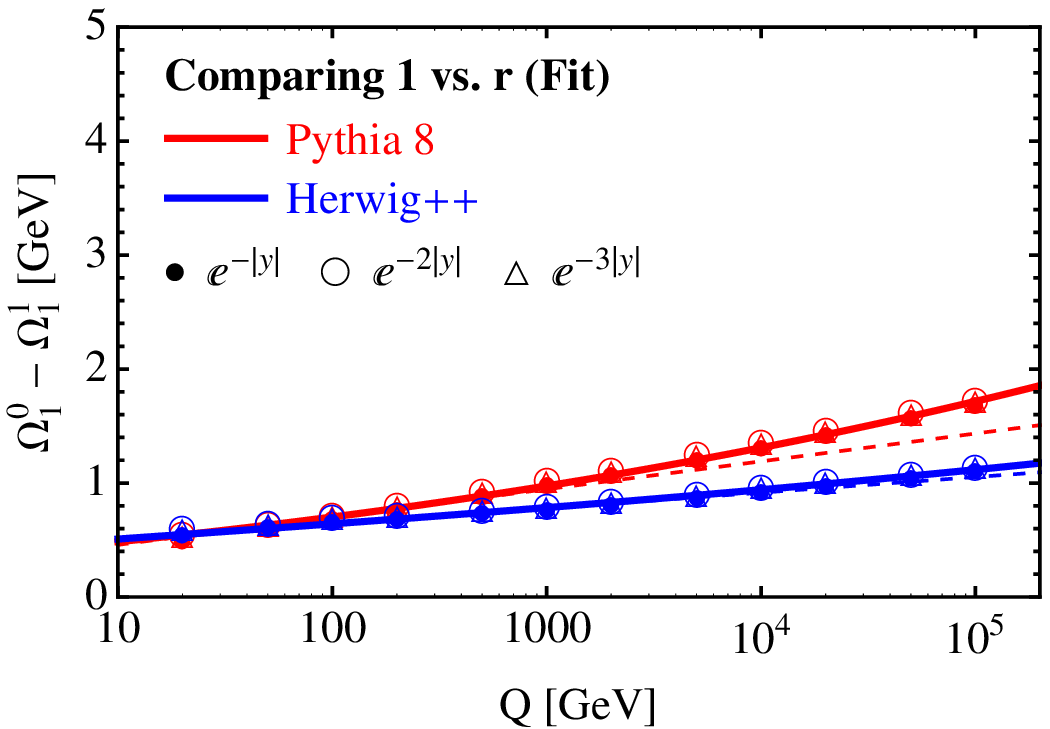}
\includegraphics[width=0.8\columnwidth]{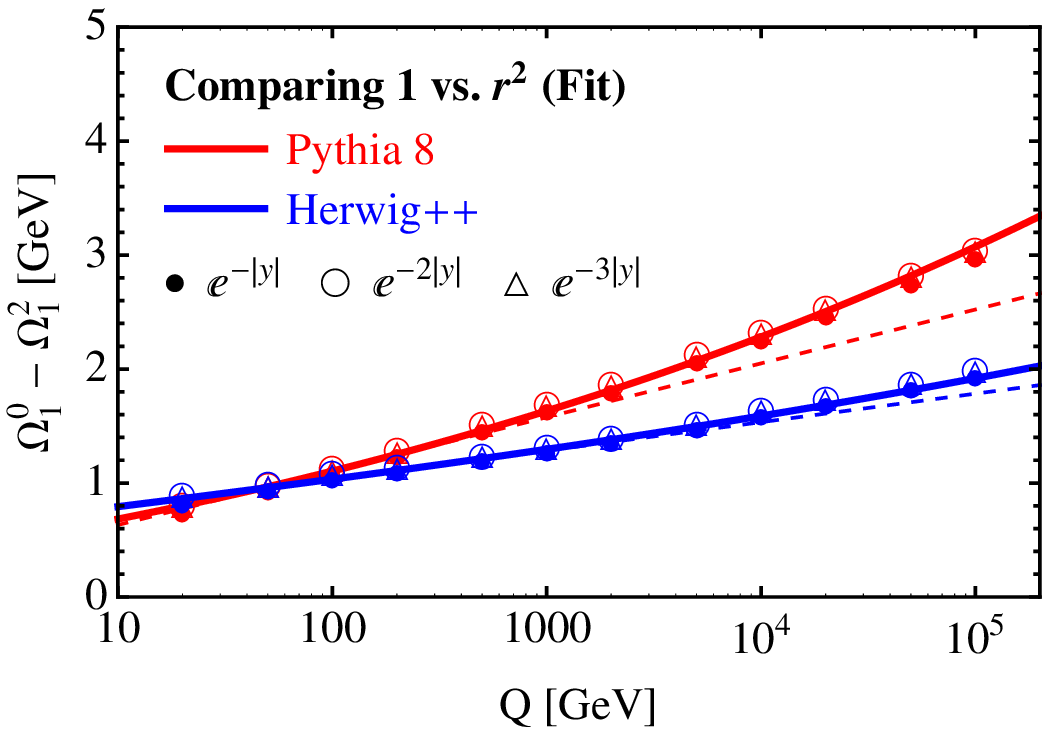}
\includegraphics[width=0.8\columnwidth]{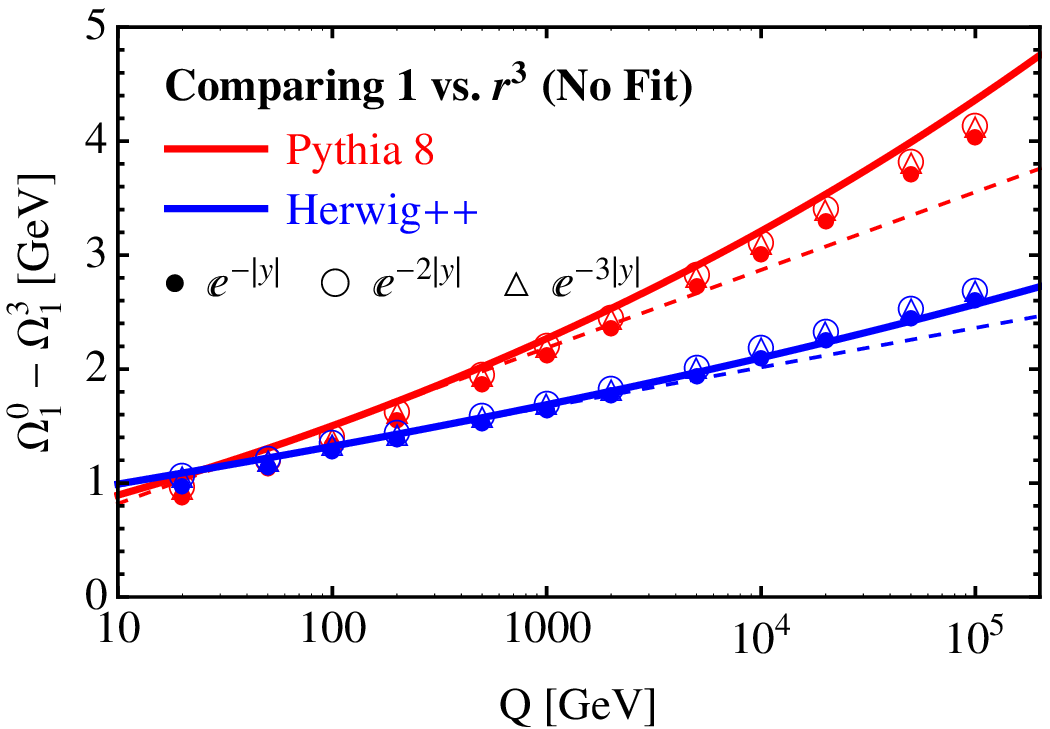}
\caption{Universal power corrections extracted from Pythia~8.162 (upper solid and
dashed lines and dots) and Herwig++ 2.6.0 (lower solid and dashed lines and
dots) for the generalized angularities $\ang{n,a}$.  We use the measured power corrections in the
first two plots to fit for $\Omega(r, \mu_0)$ at $Q = 100~\GeV$, using \Eq{eq:fullRGE}
to evolve to different $Q$ scales.  The curves in third plot are then predicted. The dashed
curves show the approximate formula in \Eq{eq:expandedrunning}.
\label{fig:mc_compare}}
\end{figure}

In \Fig{fig:mc_compare} we plot ${\Omega}_1^0 - {\Omega}_1^n$ for $a = 0,-1,-2$,
$n = 1,2,3$, and $Q$ values ranging from 20 GeV to 200 TeV.  At a fixed scale
$Q$, the power corrections are independent of $a$ with at most 5\% variations,
thus demonstrating the anticipated universality.\footnote{The leading violation
  of universality can be attributed to different matching coefficients
  $d_1^e(r)$ in \Eq{eq:C1e}.}  Note that both programs were tuned to LEP $Z$
pole and low energy data, so it is not surprising that they have the same power
corrections at $Q \sim m_Z$.  More interestingly, both programs show logarithmic
growth in $Q$ for the power correction, as expected from our results in
\Sec{subsec:onelooprun}.  Numerically, this growth is consistent with the form
$(\ln Q)^A/Q$ found in \Ref{Salam:2001bd}, and the exponent $A \simeq 4 C_A/
\beta_0 \sim 1.5$ is presumably related to the exponent in \Eq{eq:fullRGE}. A
more concrete comparison is difficult since the analysis in \Ref{Salam:2001bd}
effectively expands about $r=1$, and parametrizes the extra resulting
logarithmic singularity by a $\ln(\mu/\Lambda_{\rm QCD})$ factor that cancels an
$\alpha_s(\mu)$.

To show the importance of resummation, we fit for the functional form of
$\Omega(r, \mu_0)$.  The solid lines in \Fig{fig:mc_compare} have the full
running in \Eq{eq:fullRGE}, while the dashed lines correspond to the expansion
in \Eq{eq:expandedrunning}.  These curves were obtained following the procedure
of \Sec{sec:newbasis}, where $\Omega(r, \mu_0)$ is modeled using the three basis
functions
\be
\label{eq:basischoice}
\{1,r,(1-r)^{-1/4}\},
\ee
suitable orthonormalized.  The inclusion of $(1-r)^{-1/4}$ is needed to capture
the (integrable) peak of $\Omega(r, \mu_0)$ at $r = 1$, though other choices
$(1-r)^k$ give comparable results.  We apply the fit form at $Q= 100~\GeV$ and
use \Eq{eq:fullRGE} to determine $\Omega(r, \mu)$ over the whole $Q$
range.\footnote{As mentioned below \Eq{eq:fullRGE}, a more natural
  strategy would be to apply the fit form at $\mu_0=2~\GeV$.  Because of large
  range of scales in \Fig{fig:mc_compare} resummation is always important with
  that choice.  By using $\mu_0= (100~\GeV)/2$ we are using the same scale where
  Monte Carlos have been tuned, and we can also better highlight the difference
  between the full and expanded running.}  To show that this framework has
some predictive power, we fit $\Omega(r, \mu_0)$ using information from $n=1$
and $n=2$ over the whole $Q$ range, and then extrapolate to $n=3$.
\begin{table}[t!]
\begin{tabular}{l|cc|cc|}
&  \multicolumn{2}{c|}{$Q = 100~\GeV$} & \multicolumn{2}{c|}{$Q = 10^4~\GeV$} \\
& \,Pythia~8\, & \,Herwig++\,& \,Pythia~8\, & \,Herwig++\, \\
\hline
$\Omega_1^0 - \Omega_1^1$ & 0.7 GeV& 0.6 GeV &1.3 GeV& 0.9 GeV\\
$\Omega_1^{0,\,\ln} - \Omega_1^{1,\,\ln}$&0.9 GeV&0.5 GeV&2.4 GeV& 1.1 GeV\\
\hline
$\Omega_1^0 - \Omega_1^2$ &1.1 GeV&1.0 GeV&2.3 GeV& 1.6 GeV\\
$\Omega_1^{0,\,\ln} - \Omega_1^{2,\,\ln}$ &1.8 GeV&1.0 GeV&4.6 GeV& 2.0 GeV\\
\hline
$\Omega_1^0 - \Omega_1^3$ &1.5 GeV& 1.3 GeV &3.2 GeV& 2.1 GeV\\
$\Omega_1^{0,\,\ln} - \Omega_1^{3,\,\ln}$&2.6 GeV&1.3 GeV &6.6 GeV& 2.8 GeV
\end{tabular}
\caption{Power correction differences extracted from the fits in \Fig{fig:mc_compare}. The slope parameter $\Omega_1^{n,\,\ln}$
is defined in \Eq{eq:Olnagain}.  These values have 10\% to 20\% uncertainties from the choice of functional fit form.
\label{tab:fitcoeffs}}
\end{table}

Because we are measuring the difference $\ang{0,a}-\ang{n,a}$, we are relatively
insensitive to the functional form of $\Omega(r, \mu_0)$ near $r = 1$.  We are
effectively measuring event shapes in the universality class $g_e(r) = 1 - r^n$
where $g_e(1) = 0$, and therefore our extraction of the raw power correction
$\Omega_1^n$ has large uncertainties.  On the other hand, the extraction of
power correction differences is stable at the 10\% to 20\% level as the fit form
is adjusted, and these differences are shown in \Tab{tab:fitcoeffs} for the
basis choice in \Eq{eq:basischoice}.  

We also show the best fit values for $\Omega_1^{n,\ln}$, defined as
\begin{align} \label{eq:Olnagain}
\Omega_1^{n,\,\ln}(\mu_0) &\equiv - \!\!\int\! \df r \, \ln(1-r^2)\,
  r^n \,\Omega_1(r,\mu_0)\,,
\end{align}
which is a hadronic parameter related to the slope of the expanded running in
\Eq{eq:expandedrunning}.  We see that Pythia~8 and Herwig++ have similar power
corrections $\Omega_1^n$ at $Q = 100~\GeV$, but the slopes $\Omega_1^{n,\,\ln}$
are larger in Pythia~8, leading to larger values of $\Omega_1^n$ at $Q =
10^4~\GeV$.  This slope parameter is interesting since it provides an example of
a hadronization effect that can only be accurately determined using data at
multiple $Q$ values. Parameters of this type presumably dominate the uncertainty
one has when describing hadronization effects in high energy data using
Monte Carlo models that were only tuned at much lower energies.

Finally, we remark that one strategy to extract $\Omega_1^n$ directly from these
Monte Carlo programs would be to turn off the hadronization model and calculate
$\vev{\ang{n,a}}_\pert$ from the parton shower alone.  However, there is no
guarantee of a one-to-one map between hadronization modeling and
operator-derived power corrections.  In the context of a Monte Carlo program,
the perturbative parton shower is first evolved to the shower cutoff of order 
$1\,{\rm GeV} > \Lambda_{\rm QCD}$ before applying the hadronization model,
whereas from \Sec{sec:ope} the natural scale to evaluate the power correction is
$Qe$ (for the distribution) or $Qe_{\max}$ (for the first moment).  For this
reason, there is an ambiguous separation between perturbative parton shower
evolution and nonperturbative hadronization modeling, and there is no guarantee
that hadronization models by themselves will respect the renormalization group
evolution of \Eq{eq:fullRGE}.\footnote{For example, if we were to extract
  $\Omega_1^0$ from Pythia~8 by turning hadronization on and off, we would find
  almost no running with $Q$ (i.e. $\Omega_1^{0,\,\ln} \simeq 0$).}  It would be
interesting to understand to what extent parton shower evolution can mimic
\Eq{eq:fullRGE}, and whether there are ways to adjust hadronization models to
satisfy the renormalization group properties expected of power corrections.
Ultimately, one would want to test the power correction evolution by performing
event shape measurements at high $Q$.

\section{Conclusions}
\label{sec:conclusions}
In this paper, we revisited the important issue of power corrections for
$e^+e^-$ dijet event shapes.  By casting the leading power correction in terms
of matrix elements of a transverse velocity operator, we are able to robustly
treat the effect of hadron masses.  Depending on the measurement scheme, event
shapes fall into different universality classes that share a universal power
correction $\Omega_1^{g_e}$.  Moreover, these nonperturbative matrix elements
have perturbatively calculable anomalous dimensions, which introduce additional
dependence on the scale $Q$ of the hard collision.  

Since Monte Carlo programs play such a key role in LHC data analysis, it is
satisfying to see that both universality and $Q$-evolution are exhibited by the
hadronization models of Pythia~8 and Herwig++, albeit with different choices of
the nonperturbative matrix elements.  An interesting difference is in their
values for hadronic slope parameters that play an important role in the
extrapolation to high energies of hadronization effects that are fit at low
energies.

Our study motivates a reanalysis of $e^+ e^-$ event shape data with a more
explicit treatment of hadron mass effects. As an exercise, we have
studied the effect of including the anomalous dimension of the leading power correction
on the determination of $\alpha_s(m_Z)$ from the thrust distribution.
We have repeated the analysis of Ref.~\cite{Abbate:2010xh} at
N${}^3$LL\,+\,${\cal O}(\alpha_s^3)$, using the same
data set and procedure, but replacing the power correction to include a logarithmic slope term
\begin{align}
2\,\Omega_1 \to 2\,\Omega_1-\frac{\alpha_s(\mu_s)C_A}{\pi}\,\Omega_1^{\rm ln} \,
\log\Big(\frac{\mu_s}{2\,{\rm GeV}}\Big)\,,
\end{align}
where $\mu_s(\tau)\sim Q\tau$. With the current experimental data, $\Omega_1$
and $\Omega_1^{\rm ln}$ are highly correlated and cannot be determined
simultaneously. Therefore we plug in the estimate $\Omega_1^{\rm ln}=\pm
\,0.35\,{\rm GeV}\sim\pm\, \Omega_1$, and repeat the fit for $\alpha_s(m_Z)$ and
$\Omega_1$. We find that the effect of this term on $\alpha_s(m_Z)$ is $\pm\,
0.0005$, which is roughly half of the total uncertainty $\pm\,0.0011$ found in
Ref.~\cite{Abbate:2010xh}. For $\Omega_1$ the effect is $\mp\, 0.03\,$GeV, which
is comparable to the previous uncertainty of $\pm\,0.05\,$GeV. Accounting for
this additional source of uncertainty in quadrature changes the
total uncertainty in this analysis from $\pm 0.0011\to \pm 0.0012$ for
$\alpha_s(m_Z)$, and from $\pm 0.05\to \pm 0.06$ for $\Omega_1$.

If LEP data are successfully preserved \cite{Holzner:2009ew,Akopov:2012bm}, then
one could compare different mass schemes to better separate perturbative physics
from nonperturbative physics.  Such studies would be interesting in their own
right, but would also give additional input for tuning Monte Carlo hadronization
models.

An obvious generalization is to go beyond dijet event shapes and
consider shape functions for more than two Wilson lines.  This is relevant not
only for multijet studies at $e^+ e^-$ colliders but also for treating the beam
directions in hadron colliders like the LHC.  For example, the event shape
$N$-jettiness \cite{Stewart:2010tn} is a convenient variable to define exclusive
$N$-jet cross sections at the LHC, and its shape function involves $2+N$ Wilson
lines.  The anomalous dimension calculation for multiple Wilson lines is
technically more challenging but conceptually similar to the calculation
presented here. In particular, the same $\Eop(r,y,\hat t)$ can be considered for
this analysis (where here we add $\hat t$ in order to emphasize that rapidity
$y$ is defined with respect to the axis $\hat t$).  On the other hand, it is not
clear which aspects of universality will carry over to the multijet case, since
universality of dijet power corrections relied crucially on longitudinal boost
invariance.

Finally, our study sheds light on recent studies of jet substructure at the
LHC.  The jet shape $N$-subjettiness was introduced in
\Refs{Thaler:2010tr,Thaler:2011gf} (see also \Ref{Kim:2010uj}) as a complement
to the event shape $N$-jettiness.  The ratio of 2-subjettiness to 1-subjettiness
($\tau_{2/1}$) can be used to distinguish boosted $W/Z$ bosons from the
background of ordinary quark and gluon jets.  From \Tab{tab:classice} we see
that 2-jettiness and thrust are closely related, and \Ref{Feige:2012vc} used
this fact to perform a precision calculation of $\tau_{2/1}$ for boosted $W/Z$
bosons at the LHC by recycling the known resummation for thrust in $e^+e^- \to
\text{hadrons}$.  However, \Tab{tab:classes} shows that the original definitions
of 2-jettiness and thrust are not in the same universality class, which explains
why \Ref{Feige:2012vc} required a value of the power correction $\Phi_1 \equiv
\Omega_1^\rho$ that was roughly a factor of two bigger than the value for
$\Omega_1^\tau$ obtained in \Refs{Abbate:2010xh,Abbate:2012jh}. To a good
approximation, the power correction for boosted $W/Z$ bosons generates a shift
of the $\tau_{2/1}$ distribution by $\pi$ times
$\Omega_1^\rho$~\cite{Feige:2012vc}, making it all the more crucial to choose
the proper power correction.  Thus a proper treatment of hadron masses
and power corrections will be essential for precision jet substructure studies
at the LHC.

\begin{acknowledgments}
  We thank C.~Lee and G.~Salam for helpful conversations.
  This work was supported by the offices of Nuclear and Particle Physics of the
  U.S. Department of Energy (DOE) under grant numbers DE-FG02-94ER-40818 and
  DE-FG02-05ER-41360, and the European Community's Marie-Curie Research Networks
  under contract PITN-GA-2010-264564 (LHCphenOnet).  VM is supported by a Marie
  Curie Fellowship under contract PIOF-GA-2009-251174.  JT is supported by the
  DOE under the Early Career research program DE-FG02-11ER-41741.  VM, IS, and
  JT are also supported in part by MISTI global seed funds.  VM and IS thank the
  CERN theory group for hospitality while this work was being completed.  JT
  acknowledges the hospitality of the Aspen Center for Physics, which is
  supported by the National Science Foundation Grant No. PHY-1066293, as well as
  the hospitality of the White House.
\end{acknowledgments}

\appendix

\section{Derivation of the Transverse Velocity Operator}
\label{sec:derivation}
In this appendix we show that the transverse velocity operator $\Eop(r,
y)$ can be expressed in terms of the energy-momentum tensor as in
\Eq{eq:Eop-definition}.  Our analysis is analogous to that for $\Eop(\eta)$ in
\Ref{Bauer:2008dt}, where it was performed for scalar and spin-$1/2$
hadrons, as well as that of \Ref{Sveshnikov:1995vi}. We will carry out our
proof for scalar fields.

Consider the energy-momentum tensor of a free scalar particle with mass $m$ (we
will see below that interaction terms are suppressed):
\be \label{eq:stress-energy}
T^{\mu\nu}= \partial^\mu \phi\,\partial^\nu \phi-g^{\mu\nu}\mathcal{L}, 
\ee
where the plane wave expansion for the scalar field is
\begin{align}  \label{eq:phidef}
\phi(x) &= \int \!\dfrac{\df^3 \vec{p}}{(2\pi)^3 2E_p}
 \Big( a_{\vec p}\,e^{-i\,x \cdot p}+a_{\vec p}^\dagger\,e^{i\,x \cdot p}\Big),
\end{align}
where $E_p = \sqrt{p^2+m^2}$.
We will use the stationary phase approximation
\begin{align} \label{eq:stationary}
&\lim_{k\to\infty}\int \!\df x \, f(x) \, e^{i\,k\,g(x)} 
  \\
&~~~~=  \sqrt{\dfrac{2\pi}{k\,|g^{\prime\prime}(x_0)|}}\,f(x_0)\,
e^{i\, k\, g(x_0)}e^{i\,\frac{\pi}{4}\,
{\rm sign}[g^{\prime\prime}(x_0)] }, 
  \nn
\end{align}
where $g^{\prime}(x_0) = 0$.  For this formula to be applicable, $g(x)$ must
attain a minimum or a maximum in the range of integration.

After plugging the plane wave expansion of the free field $\phi$ in
\Eq{eq:phidef} into the energy-momentum tensor in \Eq{eq:stress-energy}, one can
perform all the angular integrations using \Eq{eq:stationary}, obtaining
\begin{align}
&\lim_{R\to\infty}R^3\hat{n}_i\,T^{0i}(R, R\,v\,\hat{n})\\
&\ = \lim_{R\to\infty}R\int\!\! \dfrac{\df p \,\df q\ p\,q^2}{4(2\pi)^4E_q\,v^2}
\Big( a_{\vec p_{\hat n}}\,a_{\vec q_{\hat n}}\,e^{iR(v p-E_p+v q-E_q)}
 \nn\\
&\quad +a_{\vec p_{\hat n}}\,a^\dagger_{\vec q_{\hat n}}\,
e^{iR(vp-E_p-vq+E_q)} +({\rm h.c.} \text{ and } p\leftrightarrow q)\Big),  \nn 
\end{align}
where $\vec p_{\hat n} = p\, {\hat n}$ and $\vec q_{\hat n} = q\, {\hat n}$.
The $p$ and $q$ integrals can be performed, again using \Eq{eq:stationary}, to yield
\begin{align}
\label{eq:massive}
&\lim_{R\to\infty}R^3\hat{n}_i\,T^{0i}(R, v\,R\,\hat{n}) \nn\\
&\ = \dfrac{1}{4(2\pi)^3}\dfrac{m^3v}{(1-v^2)^{\frac{5}{2}}}
\Big( a_{\tilde p}\,a^\dagger_{\tilde p} + a^\dagger_{\tilde p} \,a_{\tilde p} \Big),
\end{align}
where ${\tilde p} = \tfrac{mv}{\sqrt{1-v^2}}\,{\hat n}$.
Note that terms involving two creation or two annihilation operators drop out,
since they vanish as $1/R$ when integrated against any function of $v$ (in
particular our $g_e(r)$ in \eq{universality}).  At this point we can also see
why interaction terms in the energy-momentum tensor can be neglected.  Such
terms involve additional fields and therefore additional integrations when they
are expanded in plane waves.  These additional integrals can be performed using
the stationary phase approximation and vanish as $R\to \infty$ due to the
presence of additional powers of $1/R$.

After normal ordering, \Eq{eq:massive} can be written as
\begin{align}
&\lim_{R\to\infty}R^3\hat{n}_i\,T^{0i}(R, v\,R\,\hat{n}) \\
&~~~~=  \int \!\dfrac{\df^3 \vec{p}}{(2\pi)^3 2 E_p}\, a^\dagger_{p}\,a_{p}
\, \frac{E_p}{v} \, \delta\Big(v-\dfrac{|\vec{p}\,|}{E_p}\Big)
  \delta^2({\hat p}-{\hat n})\,.\nn
\end{align}
Using the fact that $E_p  =(p^\perp \!\cosh\eta) /v$ and
\begin{align}
\int_0^{2\pi} \!\! \df \phi\,\delta^2({\hat p}-{\hat n}) &= \cosh^2\!\eta
\ \delta(\eta-\eta_p)\,,
\end{align}
we obtain the operator
\begin{align}
 \Eop(v,\eta) = \frac{v^2}{\cosh^3\!\eta}  
  \lim_{R\to\infty}\! R^3 \!\! \int_0^{2\pi}\!\!\!\!\! \df \phi\: 
    \hat{n}_i\, T^{0i}(R,R\, v\,\hat{n}),
\end{align}
 which is differential in
velocity and pseudo-rapidity and satisfies:
\begin{align}
  \Eop(v, \eta)\! \ket{X} 
  & = \sum_{i \in X} p_i^\perp\delta(v-v_i)\, \delta(\eta - \eta_i) \ket{X}\,.
\end{align}
Note that if we integrate this operator over $0< v < 1$, we recover the
expression in \Ref{Sveshnikov:1995vi} for the energy flow operator $\Eop(\eta)$.

Finally, to obtain from $\Eop(v,\eta)$ the desired $\Eop(r,y)$ that satisfies
\Eq{eq:trans-mom-flow-op}, one needs to multiply by the
Jacobian factor
\begin{equation}
\dfrac{\partial(v,\eta)}{\partial(r,y)} = \dfrac{{\rm sech}^2 y}{r}\,,
\end{equation}
and include a factor $1/r$ to convert $p^\perp$ to $m^\perp$, yielding
\begin{align}
  \Eop(r,y) &= \dfrac{{\rm sech}^2 y}{r^2}\ \Eop\big(v(r,y),\eta(r,y)\big) \,,
\end{align}
which agrees with \Eq{eq:Eop-definition}.


\section{Renormalon Computation for Generic Event Shape}
\label{app:renormalon}

Here we show that the definition $\delta_e(R,\mu)
=(c_e/c_{e'})\delta_{e'}(R,\mu)$ in \eq{deltadefn} yields a perturbative
cross section $\tilde \sigma_e(x)$ in \eq{sigmay} that is independent of the
leading $\Lambda_{\rm QCD}$ renormalon when probed by a standard fermion
bubble chain. Since the renormalon cancels between the $\overline{\rm MS}$
series $\hat\sigma_e(x)$ and $\Omega_1^e(\mu)$, this implies that
$\Omega_1^e(R,\mu)$ is also free of the $\Lambda_{\rm QCD}$ renormalon.

The
$\Lambda_{\rm QCD}$ renormalon corresponds to a $u=1/2$ pole in the Borel
transform. For a function $f(\alpha_s)$ that is an infinite series in
$\alpha_s(\mu)$, the Borel transform $B[f](u)$ is obtained by replacing
\begin{align}
 \Big(\frac{\beta_0\alpha_s(\mu)}{4\pi} \Big)^{n+1} \to \frac{u^n}{n!} \,.
\end{align}
Following \Ref{Hoang:2007vb} we make use of the fact that the perturbative soft
function carries the leading renormalon, and hence carry out our computation for
$S_e^{\rm pert}(x,\mu)$ rather than the cross section
$\hat\sigma(x)$.\footnote{Note that $x$ is a dimensionless variable in
  $\hat\sigma(x)$ but is a variable with mass dimension $-1$ in $S^e_{\rm pert}(x,\mu)$.}
  Since the soft function obeys non-Abelian
exponentiation~\cite{Gatheral:1983cz,Frenkel:1984pz} it is useful to write the
perturbative scheme change in \eq{sigmay} as
\begin{align}
 \ln \tilde S_e^{\rm pert}(x,\mu) =  \ln S_e^{\rm pert}(x,\mu) - ix \,
 \delta_e(R,\mu) \,,
\end{align}
and then demonstrate that $\ln \tilde S_e^{\rm pert}(x,\mu)$ does not have a
$u=1/2$ pole.

The use of the soft function allows us to perform the bubble chain analysis for
an arbitrary event shape specified by $f_e(r,y)$ and \eq{f-definition}. To study
the first contribution to the $u=1/2$ pole, we can work in $d=4$ dimensions and
we only need to dress a single real gluon with a bubble chain. We parametrize
the gluon phase space with $\vec p_\perp$ and $y$,
\begin{align}
\dfrac{\df^{3}\vec{p}}{(2\pi)^{3}2E_p}
 =\dfrac{\df y}{4\pi}\dfrac{\df^{2} \vec{p}_\perp}{(2\pi)^{2}}\,.
\end{align}
Since the final state gluon is on-shell, we have $r=1$.  For the event shape $e$,
the Fourier transform gives
\begin{align}
  \int \!\df e\: e^{-ie x Q}\, \delta\Big(e -\frac{1}{Q} p_\perp f_e(1,y)\Big) = 
   e^{-i x\, p_\perp f_e(1,y) }\,.
\end{align}
Taking the sum of all dressed real radiation diagrams with a single gluon and
swapping $n_f\to -3\beta_0/2$, we find the Borel transform
\begin{align}
\label{eq:BorelMaster}
& B\big[\ln S_e^{\rm bubbles}(x,\mu)\big](u)
\\
  &=\frac{8C_F \big(\mu^2 e^{5/3}\big)^u}{\beta_0 \Gamma(1\!+\!u)\Gamma(1\!-\!u)}
   \int_{-\infty}^{+\infty}\!\!\!\!\!\!\!\! \df y 
  \int_0^\infty\!\!\!\!\! \df p_\perp 
  \: p_\perp^{-1-2u}\, e^{-i x\, p_\perp f_e(1,y)} 
\nn\\
 &=\frac{8C_F \big(\mu^2 e^{5/3}\big)^u}{\beta_0 \Gamma(1\!+\!u)\Gamma(1\!-\!u)}
  \int_{-\infty}^{+\infty}\!\!\!\!\!\!\!\! \df y\: f_e(1,y)^{2u} \!
  \int_0^\infty\!\!\!\! \df h\: h^{-1-2u} e^{-ixh} 
\nn\\
 &=\frac{8C_F \big(\mu^2 e^{5/3}\big)^u}{\beta_0 \Gamma(1\!+\!u)\Gamma(1\!-\!u)}
   \Gamma(-2u) (ix)^{2u} 
   \int_{-\infty}^{+\infty}\!\!\!\!\! \df y\: f_e(1,y)^{2u}
\,.
 \nn
\end{align}  
Here $\beta_0=11C_A/3 -2n_f/3$, and in the second equality we used the change of
variables $h=p_\perp f_e(1,y)$.
Expanding about $u=1/2$ and using $\int \df y\: f_e(1,y)=c_e$, we arrive at the
final expression for the $u=1/2$ pole
\begin{align} \label{eq:Sebubbles}
& B\big[\ln S_e^{\rm bubbles}(x,\mu)\big](u) 
 = c_e\frac{8C_F e^{5/6}}{\pi\beta_0(u-\frac{1}{2})}\ (ix \mu) \,.
\end{align}
Here $(ix)$ corresponds to the $\delta'(\ell)$ present in \eq{shape-function}.

Using \eq{Sebubbles} we can compute the Borel transform of the subtraction
series $\delta_{e'}(R,\mu)$ for the reference event shape $e'$, which is defined
by \eq{deltadefn}. We find
\begin{align}
  B\big[\delta_{e'}(R,\mu)\big](u) &= 
 c_{e'}\frac{8C_F e^{5/6}}{\pi\beta_0(u-\frac{1}{2})}\ \mu \,.
\end{align} 
Finally, computing the leading renormalon ambiguity in $\ln \tilde S^e_{\rm pert}$,
using \eq{deltadefn} to define $\delta_e(R,\mu)$, we find
\begin{align}
  & B\big[\ln \tilde S_e^{\rm pert}(x,\mu)\big](u) 
 \\
  &\qquad = B\big[\ln S_e^{\rm pert}(x,\mu)\big](u)  
   - ix\ B\big[\delta_{e}(R,\mu)\big](u) 
  \nn\\
  &\qquad= B\big[\ln S_e^{\rm pert}(x,\mu)\big](u)  
   - ix\ \frac{c_e}{c_{e'}}\: B\big[\delta_{e'}(R,\mu)\big](u) 
  \nn\\
  &\qquad= \frac{0}{u-\frac{1}{2}} \,, \nn
\end{align}
as promised. Note the importance of using the same scale $\mu$ for the
perturbative soft function $S_e^{\rm pert}(x,\mu)$ and its subtractions
$\delta_e(R,\mu)$.

As a final comment, we remark that the renormalon analysis in this appendix
takes $r=1$ and hence does not fully probe infrared effects that depend on
hadron masses.


\section{One-Loop Anomalous Dimension}
\label{app:anomdim}
In this appendix we provide details on the calculation which yields the anomalous
dimension formula in \eq{M1bare}.
The integrals involved in determining the one-loop anomalous dimension of
$\Omega_1(r,\mu)$ from Figs.~\ref{fig:self-energy}--\ref{fig:triple-gluon} are
somewhat different from the phase space integrals for QCD gluons attached to
eikonal lines, which occur for the leading power perturbative soft function
calculation. In particular the amplitudes are similar to those occurring in
recent two-loop soft function calculations~\cite{Kelley:2011ng,Hornig:2011iu,Monni:2011gb}, but a different
measurement is made. 

As explained in \Sec{sec:running}, one of the cut gluons
corresponds to a massive adjoint source field, and $\Eop(r,y)$ acts on this
object. We will call the momentum of this source $q^\mu$, where
$q^2=m^2\ne 0$. The remaining gluon lines are standard massless QCD gluons, and 
we will denote the momentum of virtual loop integrals by $k$, and the
momenta of real gluon radiation by $p$, where $p^2=0$. 

The phase space integral over $q$ for the source is completely fixed by the
measurement, up to one trivial angular integral for $\phi_q$ in the transverse
plane. Taking the three phase space variables to be $q^+$, $q^-$, and $\phi_q$ the former two are fixed by the $\delta$-functions from \eq{trans-mom-flow-op}:
\begin{align}\label{eq:two-deltas}
& m_q^\perp \delta(r-r_q)\,\delta(y-y_q) \\
&\quad = m_q^\perp \frac{2\,m^2r}{(1-r^2)^2}\,
\delta\Big(q^+-\frac{m e^y}{\sqrt{1-r^2}}\Big)\delta\Big(q^--\frac{m e^{-y}}{\sqrt{1-r^2}}\Big)\,,\nn
\end{align}
where $m_q^\perp=\sqrt{q_\perp^2+m^2}$, $r_q=q_\perp/m_q^\perp$, and
$y_q=1/2\ln(q^+/q^-)$. For notational convenience we define an object $\hat
\Phi(r,y)$ that contains common factors associated with the source that appear in all Feynman diagrams,
\begin{align}\label{eq:Phy-def}
&{\hat \Phi}(r,y)= \frac{16\pi\alpha_sC_F}{(2\pi)^n} \!
 \int\!\!\frac{\df^{n}{\vec q}}{2E_q}
  \ \frac{m^\perp_q\, \delta(r\!-\! r_q)\,\delta(y\!-\! y_q)}{q^+q^-}\,
\,,
\end{align}
where $n=d-1=3-2\epsilon$.  Here $\hat\Phi(r,y)$ should be considered to be an
operator that can act on additional $q^+$ and $q^-$ dependence in loop and phase
space integrals, and which replaces $q^\pm$ by the functions of $m$, $r$, $y$
occurring in the $\delta$-functions in \eq{two-deltas}.  $\hat\Phi(r,y)$ is
normalized so that acting on unity with $n=3$ the integral in \eq{Phy-def} yields
the tree level result in \eq{tree-level-c}.

The general strategy to compute the anomalous dimension is to reduce each graph
to a set of master integrals. We will always partial fraction eikonal
propagators and shift numerators to obtain a set of integrals that involve only
one $p^+$ ($k^+)$ and/or one $p^-$ ($k^-)$ in a denominator. To
regulate potential IR singularities we shift the eikonal propagators by taking
$k^\pm \to k^\pm + \Delta_{n,\bar n}$ or $p^\pm \to p^\pm +\Delta_{n,\bar
  n}$. We will treat the $\Delta_{n,\bar n}$ as infinitesimal IR regulators,
which are expanded whenever possible.

We start with graphs which involve independent emission of gluons in
\Figs{fig:abelian}{fig:non-abelian}. A useful identity between
real emission phase space and virtual integrals is
\begin{align}\label{eq:real-virtual-indentity}
I_1(A,B) = 
  &\,\tilde{\mu}^{2\epsilon}\!\!\int\!\! \frac{\df^{d-1}\vec{p}}{2|\vec{p}\,|(2\pi)^{d-1}}\frac{1}{(p^+ + A)}\frac{1}{(p^-+B)}\nn \\
= & \,i\,{\tilde\mu}^{2\epsilon}\!\!\int\!\!\frac{\df^dk}{(2\pi)^d}
  \frac{1}{(k^++A)} \frac{1}{(k^-+B)}\dfrac{1}{k^2+i0} \nn\\
= &\, \frac{1}{(4\pi)^2}\,\Gamma(\epsilon)^2\Gamma(1-\epsilon)\Big(e^{\gamma_E}\frac{\mu^2}{A\,B}\Big)^{\epsilon}\,,
\end{align}
where $\tilde \mu = \mu
e^{\gamma_E}/(4\pi)$.\footnote{\Eq{eq:real-virtual-indentity} has IR
  divergences regulated by $\epsilon$ for either $A$ or $B$ zero (strictly
  speaking UV divergences cancel IR divergences in pure dimensional
  regularization, and the integral is zero).  The equality between real and
  virtual integrals is still valid for these cases.}
Using \Eq{eq:real-virtual-indentity} it is easy to determine that the
purely Abelian terms proportional to $C_F^2$ vanish (virtual graphs cancel real
radiation graphs).  In addition in the sum of non-Abelian contributions from the
graphs in \Fig{fig:non-abelian}, there are no IR divergences regulated by
$\epsilon$ or $\Delta_{n,\bar n}$, so all $1/\epsilon$'s correspond to UV
divergences. The result for these non-Abelian independent emission contributions
is
\begin{align}
M_{\rm ie}^{\rm EFT}=&\,-8\pi\alpha_s C_A\hat{\Phi}(r,y)\,I_1(q^+,q^-)\\
= & \,-\frac{C_A\,\alpha_s}{2\pi}\hat{\Phi}(r,y)
\Big[\frac{1}{\epsilon^2}-\frac{1}{\epsilon}\,\ln\Big(\frac{q^+q^-}{\mu^2}\Big)+\frac{\pi^2}{4}\nn\\
&\,+\frac{1}{2}\ln^2\Big(\frac{q^+q^-}{\mu^2}\Big)\Big]\,.\nn
\end{align}
Note that we refer to the results of this section as effective field theory (EFT) contributions since we
are performing calculations in a theory where the soft perturbative scale
($\mu_S\sim Qe$) has been integrated out.

Graphs involving the triple gluon vertex are more involved. Here individual
virtual radiation graphs have an imaginary part, which arises from the fact that
\mbox{$q^2=m^2>0$} while all virtual particles have massless propagators.  Nevertheless
the virtual diagrams can always be paired with a complex conjugate so one only
needs the real parts. To determine the anomalous dimension for $\Omega_1(r,\mu)$,
we only need the UV-divergent terms.  The finite terms would be necessary for the
matching computation of $d_1^e(r)$ in \eq{C1e}, which is not our goal here.  We
will therefore focus on graphs which contain $1/\epsilon$ UV divergences and
$\mu$ dependence. The $\mu$-dependent terms will be needed for
\App{app:matching}.

It is straightforward to verify that the sum of double cut vacuum polarization
graphs in \Fig{fig:self-energy} does not have a $1/\epsilon$ UV divergence.
This sum of graphs also do not require $\epsilon$ to regulate IR divergences, and hence have
no explicit $\mu$ dependence. Thus they do not contribute to our calculation
here.

The remaining triple gluon vertex diagrams shown in \Fig{fig:triple-gluon}
involve single cut virtual graphs and double cut real emission graphs.  The only
required UV divergent and $\mu$-dependent loop integral is
\begin{align}
&I_2(A,q^+,m) \\
&\quad = {\rm Re}\Bigg[i\,\tilde{\mu}^{2\epsilon}\!\!
 \int\!\!\frac{\df^dk}{(2\pi)^d}\frac{1}{(k^++A)}
\dfrac{1}{k^2+i0}\dfrac{1}{(k-q)^2+i0}\Bigg]\nn\\
&\quad =  \frac{-1}{(4\pi)^2} \Gamma(\epsilon) \cos(\epsilon \pi)\Big(e^{\gamma_E}\frac{\mu^2}{m^2}\Big)^{\epsilon}
\int_0^1 \df x \,\frac{[\,x\,(1-x)\,]^{-\epsilon}}{A + x q^+}\,.\nn
\end{align}
The result is IR safe for all the cases we need ($A\neq 0$ and $A\neq -q^+$),
yielding
\begin{align}\label{eq:I2}
I_2(A,q^+,m) & = -\,\frac{1}{(4\pi)^2 q^+} 
 \Big[\frac{1}{\epsilon}- 2\ln\Big(\frac{m}{\mu}\Big)\Big]
 \ln\Big(1\!+\!\frac{q^+}{A}\Big)
\nn\\
&\qquad +\ldots\,,
\end{align}
where the $+\ldots$ refers to UV finite and $\mu$-independent terms.  
The real radiation master integral is
\begin{align}
 I_3(A,q^+,m)
= &\,\tilde{\mu}^{2\epsilon}\!\!\int\!\!
 \frac{\df^{d-1}\vec{p}}{2|\vec{p}\,|(2\pi)^{d-1}}
\frac{1}{(p^+ + A)}\frac{1}{(2p\cdot q + m^2)}\nn\\
= &\, \frac{\Gamma(\epsilon)}{(4\pi)^2}\Big(e^{\gamma_E}\frac{\mu^2}{m^2}\Big)^{\epsilon}
\int_0^\infty \df x\,\frac{[\,x\,(1+x)\,]^{-\epsilon}}{A + x q^+}\,,\nn
\end{align}
which is IR safe for all cases we need ($A\neq 0$), yielding
\begin{align}
&I_3(A,q^+,m) = \frac{1}{(4\pi)^2}\frac{1}{q^+}\Bigg\{\frac{1}{2\epsilon^2}
-\frac{1}{\epsilon}\ln\Big(\frac{m}{\mu}\Big)\\
&\qquad + \ln^2\Big(\frac{m}{\mu}\Big)
-\ln\Big(\frac{A}{q^+}\Big)\Big[\frac{1}{\epsilon}-
  2\ln\Big(\frac{m}{\mu}\Big)\Big]\Bigg\}+\ldots
 \,.\nn
\end{align}
Here all $1/\epsilon$ poles are UV divergences, and again the $+\ldots$ represent UV
finite and $\mu$-independent terms.  We can always group terms with IR
divergences as $\Delta_{n,\bar n}\to 0$ into the IR finite combination
\begin{align}
 I_{23}(q^+,m)\equiv
  I_3(\Delta,q^+,m)+\,I_2(\Delta,q^+,m) .
\end{align}
The result is independent of $\Delta$
\begin{align}
I_{23}(q^+,m^2) &= \frac{1}{(4\pi)^2}\frac{1}{q^+}\bigg[\frac{1}{2\epsilon^2}-
\frac{1}{\epsilon}\ln\Big(\frac{m}{\mu}\Big) 
+ \ln^2\Big(\frac{m}{\mu}\Big)
\nn\\
& 
\qquad + \ldots
\bigg]\,.
\end{align}
The sum of all diagrams in \Fig{fig:triple-gluon} is
\begin{align}
 M_{\rm ggg}^{\rm EFT} &= 4 \pi\alpha_s C_A\hat{\Phi}(r,y) \Big[
 q^- I_{23}(q^-,m)+ q^+ I_{23}(q^+,m) \nn\\
&\qquad + q^- I_3(q^-,q^-,m) +q^+ I_3(q^+,q^+,m) +\ldots\Big]\nn\\
&\!\!\!\!\!\!\!\!\! = \frac{C_A\,\alpha_s}{2\pi}\hat{\Phi}(r,y)\bigg[\frac{1}{\epsilon^2}
 -\frac{2}{\epsilon}\ln\Big(\frac{m}{\mu}\Big)
 + 2\ln^2\Big(\frac{m}{\mu}\Big)+\ldots\bigg].
\end{align}

Finally, adding all one-loop diagrams, the $1/\epsilon^2$ terms vanish, and we
get the following final result for the UV-divergent and $\mu$-dependent terms from the graphs in \Fig{fig:triple-gluon}:
\begin{align}\label{eq:EFT-final}
M_{\rm 1loop}^{\rm EFT} &= \hat{\Phi}(r,y) 
 \frac{\alpha_s\,C_A}{\pi}\ln\Big(\frac{q^+q^-}{m^2}\Big)
 \Big[\frac{1}{2\epsilon}-\ln\Big(\frac{m}{\mu}\Big)\Big] \nn\\
 &\qquad +\ldots
\end{align}
Due to $\hat\Phi(r,y)$, we have $\ln(q^+q^-/m^2)=-\ln(1-r^2)$, and this result
reproduces the $1/\epsilon$ term shown in \Eq{eq:M1bare}.\footnote{This
  same result can also be obtained setting $\Delta_{n,\bar n}=0$ from the start
  and using dimensional regularization for the IR divergences.
In this case we would use
\begin{align}
I_2(0,q^+,m^2) &= - I_2(-q^+,q^+,m^2)
\nn \\
&= \frac{-1}{(4\pi)^2}\frac{\Gamma(\epsilon)\Gamma(-\epsilon)\Gamma(1-\epsilon)}
 {\Gamma(1-2\epsilon)}
  \frac{1}{q^+}\cos(\epsilon \pi)
  \Big(e^{\gamma_E}\frac{\mu^2}{m^2}\Big)^{\epsilon}\nn \,,
\end{align}
and
\begin{align}
I_3(0,q^+,m) &= -I_3(q^+,q^+,m) 
 = \frac{\Gamma(2\epsilon)\Gamma(-\epsilon)}{(4\pi)^2}\frac{1}{q^+}
\Big(e^{\gamma_E}\frac{\mu^2}{m^2}\Big)^{\epsilon}
 \nn\\
& = \frac{-1}{(4\pi)^2}\frac{1}{q^+}\Bigg[\frac{1}{2\epsilon^2}
 -\frac{1}{\epsilon}\ln\Big(\frac{m}{\mu}\Big)
 + \ln^2\Big(\frac{m}{\mu}\Big) 
 +\frac{5\pi^2}{24} \Bigg]\,.\nn
\end{align}
In this case the sum of all diagrams is again IR finite and again reproduces
\eq{EFT-final}.}

The expression in \Eq{eq:EFT-final} corresponds to a bare result in the EFT. The renormalized
result ${\bar M}_{\rm 1loop}^{\rm EFT}$ is simply obtained in minimal
subtraction by canceling the UV-singularity with a counterterm
$Z(r,\mu,\epsilon)$, and will be used in \App{app:matching}. For that purpose,
we also note that the sum of independent emission and triple-gluon virtual
graphs is UV finite, so the entire contribution shown in \eq{EFT-final} comes
from the real emission diagrams.

%

\section{One-Loop Matching for Thrust}
\label{app:matching}

In this appendix we consider the one-loop matching computation that determines
$C_1^e(\ell,r,\mu)$ in \eq{shape-function2} for the case where $e$ is thrust.
Only $\mu$-dependent terms will be considered since our goal is to see how the
function $\df/\df\ell[ 1/\mu (\mu/\ell)_+ ]$ arises from the matching computation.

A matching computation is performed by considering the difference of
renormalized full theory and renormalized effective theory matrix elements,
which are calculated with precisely the same infrared regulator(s). For our
computation the full theory corresponds to matrix elements of the soft function. For
an event shape $e$, it is
\begin{align} \label{eq:Sel}
S_e(\ell) & = \langle\,0\,|\, \overline{Y}_{\bar n}^\dagger Y_n^\dagger
    \delta(\ell - Q\hat e) Y_n \overline{Y}_{\bar n}\,|\,0\,\rangle \\
  & = \sum_X \langle\,0\,|\, \overline{Y}_{\bar n}^\dagger Y_n^\dagger
   \,|\,X\,\rangle  \delta[\ell - Qe(X)] 
   \langle\,X\,|\,Y_n \overline{Y}_{\bar n}\,|\,0\,\rangle
   \nn ,
\end{align}
where the sum is over all possible intermediate states and includes also
integrals over phase space. The EFT for this computation corresponds to the
field theory obtained by integrating out the scale $\ell$, and involves matrix
elements like $\Omega_1(r)$ in \eq{universalitya}. In this language the
anomalous dimension calculation in \App{app:anomdim} corresponded to finding the UV
counterterm for the $\Omega_1(r)$ matrix element in the EFT. For a matching
computation that only considers \mbox{$\mu$-dependent} terms, the required renormalized
EFT matrix element corresponds to the $\ln(m/\mu)$ term in \eq{EFT-final}. This
result is independent of the event shape $e$, in contrast to the $d_1^e(r)$
term in \Eq{eq:C1e} which is event shape specific.

Here we perform a computation of the corresponding full theory matrix
elements in \eq{Sel} with the same IR regulators, which includes the use of the
source $J^{\mu A}$ field of momentum $q^\mu$ where $q^2=m^2$. Since some parts
of this computation depend on the choice of $e$, we will restrict ourselves to a
computation for thrust. We will use the same notation as \App{app:anomdim} for
loop and phase space integrals. Unlike the EFT, the full theory results involve
hierarchical scales, $\ell \gg q^\mu \sim m$, and hence the final full theory
result must be expanded before subtracting the EFT result.

At the order of our calculation, we can split the $\sum_X$ in \eq{Sel} into terms
with no source term (which gives rise to the purely perturbative soft function),
and terms with one source term. In general the expansion $\ell \gg q$ should
only be performed after carrying out the full theory loop and phase space
integrals, but in cases where the expansion and integration commute we can do
them in either order. One example where this is useful is in the measurement
function $\delta(\ell-Qe(X))$.  Denoting by $e(q)$ the contribution from the
source, and $e(p_i)$ the contribution from all other real radiation gluons we
can expand $\ell\sim Q e(p_i)\gg Q e(q)$. Keeping the
first two terms only gives
\begin{align} \label{eq:expndel}
 & \delta[\ell-Qe(p_i)-Qe(q)] 
 \\
 &\qquad =\delta[\ell-Q e(p_i)]  -Qe(q) \,\delta'[\ell-Q e(p_i)]  \,. \nn
\end{align}
For the corresponding terms in $S_e(\ell)$ this yields
\begin{align} \label{eq:Selexpn}
S_e(\ell) &= S^{\rm pert}_e(e) - \frac{\df}{\df \ell}
      \sum_{\{{p_i}\},{q}}  Q e(q)\: \delta[\ell- Qe(p_i)] \\
  &\qquad  \times
  \langle\,0\,|\, \overline{Y}_{\bar n}^\dagger Y_n^\dagger\,|\,p_i,q\,\rangle
 \langle\,p_i,q\,|\,Y_n \overline{Y}_{\bar n}\,|\,0\,\rangle \nn
  \,.
\end{align}
The first term corresponds to the leading power perturbative soft function and
the second term provides the full theory contribution to the matching we are
interested in. The analog of \eq{expndel} for the EFT computation is
\begin{align}
 \delta[\ell -Q e(p_i)-Qe(q)]=
  \delta(\ell)-\delta'(\ell)\,[Qe(q)+Qe(p_i)] \,,
\end{align}
where the term $Qe(p_i)$ is scaleless and vanishes.  From this result we see
that the EFT contribution to the matching is proportional to $-\delta'(\ell)$.

It is easy to check that the matching is simple for an Abelian theory. Due to
the exponentiation and factorization properties of the Abelian eikonal matrix
elements,  we obtain
\begin{align}\label{eq:match-abelian}
& -\!\sum \langle 0 |\, \overline{Y}_{\bar n}^\dagger Y_n^\dagger\,
  |\,p_i,q\,\rangle
  \delta'[\ell -Q e(p_i)]\,Qe(q) \langle\,p_i,q\,
  |\,Y_n \overline{Y}_{\bar n}\,|0\rangle
  \nn\\
&\qquad  = - \frac{\df}{\df \ell}S^{\rm pert}_e(\ell) \langle 0
   |\, \overline{Y}_{\bar n}^\dagger Y_n^\dagger
   (Q\hat e) Y_n \overline{Y}_{\bar n}\,| 0 \rangle
  \nn\\
&\qquad = - \frac{\df}{\df \ell}S^{\rm pert}_e(\ell)\ \Omega_1^e
  \,.
\end{align}
This result holds even if we consider including more than one source term. In
\Ref{Abbate:2010xh} it was assumed that \Eq{eq:match-abelian} also
encoded all non-Abelian contributions. While these non-Abelian contributions are
indeed present, \eq{C1e} implies that in general there are additional
non-Abelian corrections from the $+$-function and $d_1^e(r)$ terms. The flaw in
the argument in Appendix B of \Ref{Abbate:2010xh} is that the dimension-1
operator $\Eop(\eta)$ is not unique, since there exists an entire family
of operators $\Eop(y,r)$ parametrized by $r$.
 
To carry out the full non-Abelian calculation for the second term in
\eq{Selexpn}, we use \eqs{trans-mom-flow-op}{e} to decompose $Qe(q)$ and write the full amplitude
as
\begin{align}
  A^{\rm Full}_\tau(\ell)
  = \int \!\df r\: \df y\: f_\tau(r,y) M_\tau^{\rm Full}(\ell,r,y) \,.
\end{align}
Results for $M_\tau^{\rm Full}(\ell,r,y)$ can then be compared directly to the
analogous results for $-M^{\rm EFT}(r,y)\delta'(\ell)$ obtained with various
$M^{\rm EFT}$ results from \App{app:anomdim}.  Here $M^{\rm Full}_\tau$ still
depends on the thrust event shape because it contains $\delta(\ell-Qe(p_i))$.

The computation with one source and no additional gluons is very simple, and we
find 
\begin{align}
M_{\rm tree}^{\rm Full}=-\,\hat{\Phi}(r,y)\,\delta'(\ell)\,,
\end{align}
where $\hat\Phi(r,y)$ is given in \eq{Phy-def}.
Next consider the computation for thrust at one loop. One performs the
master-integral decomposition in the same way as for the anomalous dimension
computation. The sum of all full theory virtual diagrams is UV finite and
$\mu$-independent (once one performs the usual QCD renormalization).  In fact
the sum of diagrams involving a virtual gluon is identical in the full and EFT
computations, and hence these contributions cancel when subtracting to determine
the matching. 

Thus we only need master integrals involving full theory diagrams with real
radiation to complete the calculation. For these contributions only two new
master integrals are required to compute the $\mu$-dependent pieces.  For the
thrust measurement on the real radiation gluon we will use the short-hand
notation
\begin{align}
 {\cal M}_\tau(\ell,p^\pm) 
  \equiv -\frac{\df}{\df \ell}\Big[ & \delta(\ell-p^+)\theta(p^- \!-\! p^+)
  \\
  & +\delta(\ell-p^-)\theta(p^+ \!-\! p^-) \Big] ,\nn
\end{align}
and to expand the master integrals with $q\ll \ell$ we will use the identity
\begin{align}
\frac{x^\epsilon}{x+\delta}&=\Big[\frac{1}{x}\Big]_+ -\delta(x)\,\ln(\delta)+
\epsilon\,\Big[\frac{\ln(x)}{x}\Big]_+\\
&\qquad -\epsilon\,\delta(x)
  \Big(\frac{1}{2}\,\ln^2\delta+\frac{\pi^2}{6}\Big)
  +{\mathcal O}(\epsilon^2,\delta)\,. \nn
\end{align}

The first master integral shows up in the independent emission diagrams of
\Fig{fig:non-abelian} and when expanded for $q\ll \ell$ gives
\begin{align}
&\tilde{\mu}^{2\epsilon}\!\!\int\!\! 
  \frac{\df^{d-1}\vec{p}}{2|\vec{p}\,|(2\pi)^{d-1}}
  \frac{1}{p^+ + A}\frac{1}{p^- + B}\, {\cal M}_\tau(\ell,p^\pm)
  \\
&\quad = \frac{-1}{(4\pi)^2}\frac{\df}{\df\ell}\Bigg\{\,\frac{1}{\epsilon}
  \bigg[\frac{2}{\mu}\,\Big(\frac{\mu}{\ell}\Big)_+
-\delta(\ell)\ln\Big(\frac{A\,B}{\mu^2}\Big)\bigg]
\nn\\
&\quad -\frac{4}{\mu}\Big[\frac{\mu}{\ell}\ln\Big(\frac{\mu}{\ell}\Big)\Big]_+
 \!
+\delta(\ell)\Big[\ln^2\Big(\frac{A}{\mu}\Big)\!+\ln^2\Big(\frac{B}{\mu}\Big)
 \!+\! \frac{2\pi^2}{3}\Big]\Bigg\}
 \,.\nn
\end{align}
The second master integral appears in the double cut graphs involving the triple
gluon vertex in \Fig{fig:triple-gluon}, and when expanded for $q\ll \ell$
reads
\begin{align}
&\tilde{\mu}^{2\epsilon}\!\!\int\!\! \frac{\df^{d-1}\vec{p}}{2|\vec{p}\,|(2\pi)^{d-1}}
\frac{1}{p^+ + A}\frac{1}{2p\cdot q+m^2}\, {\cal M}_\tau(\ell,p^\pm)=
  \nn\\
& \frac{-1}{(4\pi)^2}\frac{1}{q^+}\frac{\df}{\df\ell}
  \Bigg\{\bigg[\frac{1}{\epsilon}
  -\ln\Big(\frac{m^2}{q^{+\,2}}\Big)\bigg]
 \bigg[\frac{1}{\mu}\Big(\frac{\mu}{\ell}\Big)_+ 
 \!\! -\delta(\ell)\ln\Big(\frac{A}{\mu}\Big)\bigg]
  \nn\\
&-\frac{2}{\mu}\Big[\frac{\mu}{\ell}\ln\Big(\frac{\mu}{\ell}\Big)\Big]_+
+ \delta(\ell)\ln^2\Big(\frac{A}{\mu}\Big)\Bigg\}+\ldots\,,
\end{align}
where the omitted terms in the $+\ldots$ are $\mu$-independent and
\mbox{UV-finite}.

Adding up all the $\mu$-dependent full theory (real radiation) diagrams we find
\begin{align}\label{eq:full-theory-final}
M_{\rm 1loop}^{\rm Full}(r,y) &=
 \hat{\Phi}(r,y)
 \frac{C_A\alpha_s}{\pi} \ln(1-r^2)
\bigg\{\frac{\df}{\df\ell} \frac{1}{\mu}\Big[\frac{\mu}{\ell}\Big]_+
  \nn \\
&\qquad -\delta'(\ell)\ln\Big(\frac{m}{\mu}\Big)\bigg\} +\ldots
\end{align}
Note that $\mu\, \df/\df\mu \, M_{\rm 1loop}^{\rm Full}(r,y)=0$ since the two
contributions cancel each other. This is consistent with the fact that there was
no UV or IR divergence regulated by $\epsilon$. We still refer to them as
$\mu$-dependent terms since they have different functional dependence on $\ell$.
In \eq{full-theory-final} $\mu$ is simply a place holder scale for splitting the
result into $+$-function and $\delta$-function terms.  For the corresponding EFT
result, using \eq{EFT-final} to get the sum of renormalized diagrams, we have
\begin{align} \label{eq:M1eft}
 -\delta'(\ell) \bar M_{\rm 1loop}^{\rm EFT} 
 &=-  \delta'(\ell) \hat{\Phi}(r,y)
 \frac{C_A\alpha_s}{\pi} \ln(1-r^2) \ln\Big(\frac{m}{\mu}\Big)
  \nn\\
 &\qquad  +\ldots
\end{align}
When we subtract \eq{M1eft} from \eq{full-theory-final} we are left with only
the $\mu$-dependent $+$-function term.  Identifying $\int \df r\,  \df y\, 
f_\tau(r,y) \hat\Phi(r,y)=\Omega_1^\tau$, this reproduces the $\mu$-dependent
term in the matching result given in \Eq{eq:C1e}.

\bibliography{thrust3}
\end{document}